\documentclass{jfm}

\usepackage{graphicx}
\usepackage{epstopdf, epsfig}
\usepackage{amsmath}
\usepackage{amssymb}
\usepackage{fancybox} 
\usepackage{natbib}
\usepackage{color}
\usepackage{subcaption}
\usepackage{tikz}

\definecolor{C0}{RGB}{255,0,0}
\definecolor{C1}{RGB}{0, 127, 0}
\definecolor{C2}{RGB}{31,119,180}
\definecolor{C3}{RGB}{255,127,14}
\definecolor{C4}{RGB}{148,103,189}
\definecolor{F1}{RGB}{190,0,0}
\definecolor{F2}{RGB}{0,192,0}
\definecolor{F3}{RGB}{0,0,242}

\begin{document}

\newtheorem{lemma}{Lemma}
\newtheorem{corollary}{Corollary}

\newcommand{\dd}[2]{\displaystyle\frac{\partial#1}{\partial#2}}
\newcommand{\moy}[1]{\left\langle #1 \right\rangle}
\newcommand{\vecnum}[1]{\left\{#1\right\}}
\newcommand{\matnum}[1]{\left[#1\right]}

\shorttitle{Multi-phase flow modelling of grain-size segregation}
\shortauthor{H. Rousseau et al.}

\title{Bridging the gap between particle-scale forces and continuum modelling of size segregation: application to bedload transport}

\author
{
	Hugo Rousseau\aff{1}
	\corresp{\email{hugo.rousseau@inrae.fr}},
	Remi Chassagne\aff{1},
	Julien Chauchat\aff{2},
	Raphael Maurin\aff{3},
	\and 
	Philippe Frey\aff{1}
}

\affiliation
{
	\aff{1}
	Univ. Grenoble Alpes, INRAE, UR ETNA, 38000 Grenoble, France
	\aff{2}
	Univ. Grenoble Alpes, LEGI, CNRS, UMR 5519 - Grenoble, France
	\aff{3}
	IMFT, Univ. Toulouse, CNRS-Toulouse, France
}
	
\maketitle
	
\begin{abstract}
	Gravity-driven size segregation is important in mountain streams where a wide range
	of grain sizes are transported as bedload. More particularly, vertical size segregation is a multi-scale process that originates in interactions at the scale of particles with important morphological consequences on the reach scale. 
	To address this issue, a volume-averaged multi-phase flow model for immersed bidisperse granular flows was developed based on an interparticle segregation force \citep{GuillardScalinglawssegregation2016} and a granular Stokesian drag force \citep{TripathiDensitydifferencedrivensegregation2013}. An advection-diffusion model was derived from this model yielding parametrisations for the advection and diffusion coefficients based on the interparticle interactions. This approach makes it possible to bridge the gap between grain-scale physics and continuum modelling.
	Both models were successfully tested against existing Discrete Element Model (DEM) simulations of size segregation in bedload transport \citep{Chassagne2020}. Through a detailed investigation of the granular forces, it is demonstrated that the observed scaling of the advection and diffusion coefficients with the inertial number can be explained by the granular drag force dependency on the viscosity. 
	The drag coefficient was shown to be linearly dependent on the small particle concentration. The scaling relationship of the segregation force with the friction coefficient is confirmed and additional non-trivial dependencies including the inertial number and small particle concentration are identified.
	Lastly, adding a size ratio dependency in the segregation force perfectly reproduces the DEM results for a large range of small particle concentrations and size-ratios.  
\end{abstract}	
	
\section{Introduction}
Bedload transport, the coarser sediment load transported by the water flow in close contact with the mobile river bed, is a major process that shapes the Earth surface with consequences for public safety, water resources, territorial development and fluvial ecology. In mountain streams with steep slopes, large quantities of a wide range of grain sizes are transported leading to grain size sorting more generally named size segregation. Size segregation remains a poorly understood phenomenon \citep{GrayParticleSegregationDense2018} impairing our ability to model the interplay between sediment transport rates and channel morphological evolution such as armouring \citep{BathurstEffectCoarseSurface2007}, bedload sheets \citep{VendittiMobilizationcoarsesurface2010, Bacchieffectskineticsorting2014}, patching \citep{NelsonBedtopographydevelopment2010} or downstream fining \citep{PaolaDownstreamFiningSelective1992}. The physics of granular media has been advocated to address segregation at the granular scale and understand geomorphological evolution \citep{FreyHowRiverBeds2009, FreyBedloadgranularphenomenon2011}.
Size segregation largely originates from local interparticle interactions but has huge consequences on the particle size repartition both in the downward and streamwise directions over a much larger scale, potentially affecting sediment mobility and the entire channel geomorphological equilibrium \citep{Gilbert1914, FergusonReconstructingsedimentpulse2015, DudillInfiltrationfinesediment2017, DudillIntroducingFinerGrains2018}.
While investigating segregation at the granular scale (usually with discrete methods) is invaluable \citep{hill_tan_2014, FerdowsiRiverbedarmouringgranular2017, Chassagne2020}, it is also necessary to consider continuum modelling to improve our theoretical understanding and to provide predictions at larger scales. The focus of this paper is therefore to bridge the gap between the granular scale processes and continuum modelling, by determining closures based on local granular mechanisms.\\

This contribution focuses on vertical size segregation processes due to kinetic sieving and associated squeeze expulsion \citep{SavageParticlesizesegregation1988,GrayParticleSegregationDense2018}. The moving particles act as a random fluctuating sieve, in which small particles are more likely to percolate under the action of gravity than larger particles. This downward movement is balanced by upward squeeze expulsion which equally applies on small and large particles resulting in a net downward motion of the small particles. The combination of both processes is called gravity driven segregation (Gray 2018) and is the dominant mechanism in bedload transport. Beyond the few studies made on size segregation in bedload transport \citep{HergaultImageprocessingstudy2010, FerdowsiRiverbedarmouringgranular2017, FreyExperimentsgrainsize2020,Chassagne2020}, these processes have been studied experimentally and numerically in many granular flows such as dry granular avalanches \citep{SavageParticlesizesegregation1988, DolguninSegregationmodelingparticle1995, WiederseinerExperimentalinvestigationsegregating2011, JonesAsymmetricconcentrationdependence2018, Thorntonthreephasemixturetheory2006,GuillardScalinglawssegregation2016}, shear cells \citep{GolickMixingsegregationrates2009, vanderVaartUnderlyingAsymmetryParticle2015} or annular rotating drums \citep{ThomasReverseintermediatesegregation2000}.\\

While particularly complex segregation phenomena were observed \citep{ThomasReverseintermediatesegregation2000}, size segregation has been found to be mainly related to the forcing, the size-ratio and the fine particle volume fraction. \citet{SavageParticlesizesegregation1988} predicted from dimensional analysis that the shear rate $\dot{\gamma}^p$ should be the controlling parameter for size segregation. Indeed, when a granular medium is sheared, a layer of particles moves relatively faster than the one beneath, allowing particles to find gaps in which to fall by gravity. This theory agrees with experimental bidisperse flow down inclined planes \citep{SavageParticlesizesegregation1988}. In more recent works, \citet{GolickMixingsegregationrates2009} with shear cell experiments, and \citet{FryEffectpressuresegregation2018} with shear cell Discrete Element Model (DEM) simulations, evidenced the effect of granular pressure, observing less efficient segregation when increasing the pressure. \citet{GrayParticleSegregationDense2018} suggested size segregation to depend on the inertial number, classically used to describe granular rheology \citep{GDRMiDidensegranularflows2004,DaCruzRheophysicsdensegranular2005},
\begin{equation}
I = \dfrac{d_l \dot{\gamma}^p}{\sqrt{P^p/\rho^p}},
\label{inertial_number}
\end{equation}
where $d_l$ is the large particle diameter, $\dot{\gamma}^p$ is the granular shear rate, $P^p$ is the granular pressure, and $\rho^p$ is the particle density. DEM simulations of dry granular flows \citep{FryEffectpressuresegregation2018} and turbulent bedload transport \citep{Chassagne2020} have shown that the segregation velocity indeed scales with the inertial number to a power $0.845\pm0.05$ from quasi-static to dense granular flow regimes.\\
Not surprisingly, a number of studies have also found that the segregation depends on the particle size ratio in the kinetic sieving regime. As kinetic sieving is related to the gaps created by shearing, it appears logical to be related to the size ratio. However, while \citet{Chassagne2020} have found that segregation increases monotonically with the size ratio in quasi-static regimes ($I<10^{-3}-10^{-2}$) for size ratio up to 3, \citet{GolickMixingsegregationrates2009} and \citet{GuillardScalinglawssegregation2016} found that it experiences a maximum efficiency for a size ratio of two, $r=2$, for more dynamic granular regimes ($10^{-3}<I<1$). Yet, there is still no satisfying theory that explains this difference.\\
Similarly to the hindrance function for the fluid drag force on a particle, size segregation has also been observed to depend on the concentration of fine particles. Indeed, studies indicate that the efficiency of the segregation process is linked to concentration in small (or large) particles in the granular sample \citep{FanModellingsizesegregation2014, vanderVaartUnderlyingAsymmetryParticle2015, JonesAsymmetricconcentrationdependence2018}.\\

Size segregation can be analysed both from a particle-scale mechanistic point of view, or a continuum one. On the one hand, considering a large particle in a bath of small particles, size segregation can be seen as a force destabilizing the large particle and leading to a migration with respect to the small particles. A number of authors have adopted this approach and have shown that a particle experiences different kind of forces linked to size segregation \citep{DingDragInducedLift2011, TripathiDensitydifferencedrivensegregation2013, GuillardScalinglawssegregation2016, StaronRisingdynamicslift2018, vanderVaartSegregationlargeparticles2018}. The forces can be decomposed into a component that drive the segregation and a resulting resisting component linked to the relative motion of the large particle with respect to the small. These two different forces have been isolated by  \citet{GuillardScalinglawssegregation2016} and \citet{TripathiDensitydifferencedrivensegregation2013}. To assess the segregation forces due to the particle size differences, \citet{GuillardScalinglawssegregation2016} performed 2D DEM simulations of a large disk placed in a bed of small disks in the simple shear flow configuration. Maintaining the large disk at a given position with a virtual spring, they were able to assess the vertical segregation force applied by the small particles to the large one, $f_{seg}$ (see figure \ref{single_in_smalls}), without generating a resisting force due to the particle motion. \citet{GuillardScalinglawssegregation2016} found that the vertical segregation force has two contributions: one proportionnal to the pressure gradient $\partial P^p / \partial z$ arising from the enduring contact between particles; the other proportional to the granular shear stress gradient $\partial{|\tau^p|}/\partial{z}$:
\begin{equation}
f_{seg} = V_l \Big(\mathcal{F}(\mu,r) \dd{P^p}{z} + \mathcal{G}(\mu,r) \dfrac{\partial{|\tau^p|}}{\partial{z}} \Big),
\label{segregationForce_intro}
\end{equation}
where $V_l = \pi d_l^{3}/6$ is the volume of the intruder and $\mathcal{F}$ and $\mathcal{G}$ are empirical functions depending on the friction coefficient $\mu = |\tau^p|/P^p$ and on the size ratio $ r = d_l/d_s$ between the intruder and the surrounding small particles. \citet{GuillardScalinglawssegregation2016} studied the dependency on both parameters but only provided a dependency with $\mu$ as
\begin{equation}
\mathcal{F}(\mu) = 2.4 + 0.73 \, e^{ \textstyle -\frac{(\mu-\mu_c)}{0.051}}, \quad \text{and} \quad \mathcal{G}(\mu) = - \left(2+ 5.5 \, e^{ \textstyle -\frac{(\mu-\mu_c)}{0.076}}\right),
\label{function_Guillard}
\end{equation}
where $\mu_c$ is the critical friction coefficient defining the threshold of movement. \\
\citet{TripathiDensitydifferencedrivensegregation2013} performed 3D DEM simulations of a settling heavy sphere in a bed of lighter spheres, during a steady dry granular flow on an inclined plan. This density segregation setup generates a relative motion between the heavy sphere and the lighter ones, without generating segregation forces due to size ratio. By analogy with classical hydrodynamics, light particles playing the role of an ambiant fluid, the authors showed that the interaction force could be modeled with a Stokesian form of a solid drag force
\begin{equation}
f^{p}_{d} = c(\Phi) \pi \eta^{p} d_l v,
\label{solid_drag_intro}
\end{equation}
where $v$ is the settling velocity of the heavier particle, $c(\Phi)$ is a drag coefficient depending on the local solid volume fraction $\Phi$, and $\eta^p = |\tau^p|/|\dot{\gamma}^p|$ is the viscosity of the granular medium considered as a non-Newtonian fluid. \citet{TripathiDensitydifferencedrivensegregation2013} suggested that $c(\Phi)$ depends on the local volume fraction $\Phi$ but still remains of the same order as the value observed for a Stokes law in Newtonian fluids, i.e. $c=3$. These two forces allow one to understand particles migration locally, and to relate the segregation behaviour of particles to the local characteristics of the granular flow.\\

By contrast, addressing the effect of size segregation processes at the large scale requires a different approach that disregard the particles. Such an approach has been extensively developed in the last few years focusing on a description of segregation as an advection-diffusion model for the percolation of small particles
\citep{DolguninDevelopmentmodelsegregation1998,Thorntonthreephasemixturetheory2006, GrayParticlesizesegregationdiffusive2006,vanderVaartUnderlyingAsymmetryParticle2015, FerdowsiRiverbedarmouringgranular2017,GrayParticleSegregationDense2018, CaiDiffusionsizebidisperse2019}:
\begin{equation}
\dfrac{\partial \phi^s}{\partial t} - \dfrac{\partial}{\partial z}\big(\phi^s w_s) =   \dfrac{\partial}{\partial z}\big(D \dfrac{\partial \phi^s}{\partial z}  \big),
\label{adv-diff_intro}
\end{equation}
where $t$ denotes for time, $z$ for the vertical axis, $\phi^s$ and $\phi^l$ are the small and large particle concentration and sum to unity (with $\phi^s+\phi^l = 1$), $w_s$ is the advection velocity of segregation and $D$ is the diffusion coefficient. This equation is characterised by the segregation flux $\phi^s w^s$, and the advective velocity $w^s$, which encompass the physical dependencies of size segregation discussed previously. The advective velocity should therefore have a dependence on the local concentration which is classically taken as proportional to the large particle concentration, $w_s=\phi^l S_r$ \citep{BridgwaterParticlemixingsegregation1985, SavageParticlesizesegregation1988, DolguninSegregationmodelingparticle1995, Graytheoryparticlesize2005, GajjarAsymmetricfluxmodels2014, FanModellingsizesegregation2014, JonesAsymmetricconcentrationdependence2018}. $S_r$ is called the advection coefficient and it has been usually taken as an empirical constant for a given application, or determined from semi-empirical analysis. Based on a dimensional analysis and DEM simulations, \citet{Chassagne2020} showed that it should depend on both the inertial number and the size ratio. The diffusion coefficient $D$ models the diffusive remixing of small particles into large particles. Contrary to the advection coefficient, the diffusion coefficient has received less attention in the literature. It has been suggested that it should depend on the volume fraction \citep{CaiDiffusionsizebidisperse2019} and on the inertial number \citep{Chassagne2020}.

Developing a three-phase continuum mixture theory to model a bi-disperse combination of large and small particles with an interstitial passive fluid, \citet{Thorntonthreephasemixturetheory2006} and \citet{GrayParticlesizesegregationdiffusive2006} were able to analytically derive the advection-diffusion model (\ref{adv-diff_intro}). This derivation represented an important step in the understanding of the physical processes at work in size segregation since the advection and diffusion coefficients of the advection-diffusion equation were linked to the particle-scale interactions. In particular, the derivation is based on the assumption that the size segregation directly takes its origin in the heterogeneous distribution of the granular pressure between small and large particles. However, the form of the interaction forces between large and small particles have been postulated without support from independent physical evidence.\\

This literature review evidences the absence of direct link between the continuum modelling of grain-size segregation and the local segregation forces experienced by a grain. In this context, the aim of the present paper is to bridge the gap between the granular scale approach and the continuum modelling. Based on the particle-scale forces proposed by \citet{GuillardScalinglawssegregation2016} and \citet{TripathiDensitydifferencedrivensegregation2013}, a volume-averaging approach \citep{JacksonLocallyaveragedequations1997, Jacksondynamicsfluidizedparticles2000} is adopted here to derive a multi-phase continuum model from granular-scale forces. In addition to the novelty of the developed approach, the derivation proposed by \citet{Thorntonthreephasemixturetheory2006} is used to express the advection-diffusion equation from the new multi-phase flow model, providing improved formulations of the advection and diffusion coefficients that contain the particle-scale granular dependencies. In order to test the proposed models, the bidisperse turbulent bedload transport configurations investigated in \citet{Chassagne2020} are used for comparison. The DEM simulations performed by the authors give a good reference in which granular-scale processes are explicitly resolved. In addition it makes it possible to focus on size segregation by providing an input for the granular rheology. As the study of \citet{Chassagne2020} focused on the quasi-static part of the bed in turbulent bedload transport, the comparison will be mainly performed in this regime. Since the fluid turbulence can be neglected in this regime \citep{MaurinDensegranularflow2016}, it will not be taken into account in the derivation of the equations.\\

The paper is organised as follows: first, the forces acting at the granular scale for a single large intruder in an immersed sheared granular flow are discussed. Then, in section \ref{multi_phase_section}, the multi-phase flow model is derived by volume-averaging and the associated advection-diffusion equation is derived in section \ref{section_adv_diff}. Finally, both models are compared to the DEM simulations (section \ref{results}) and ways to improve the closures are discussed in section~\ref{discussion_part}, including the influence of the size ratio.

\section{A large intruder in a bath of small particles}
\label{largeIntruderInSmalls}
As a first step, the force balance applied on a single large grain in an immersed granular medium made of smaller particles is presented. This Lagrangian equation of motion for the large intruder is then made dimensionless using classical scalings for granular flows, with the large particle diameter as the length scale. An order of magnitude analysis makes it possible to discriminate the most important forces for bedload transport application.
\subsection{Force balance on the large intruder} 
\begin{figure}
	\centering{\includegraphics[scale=0.7]{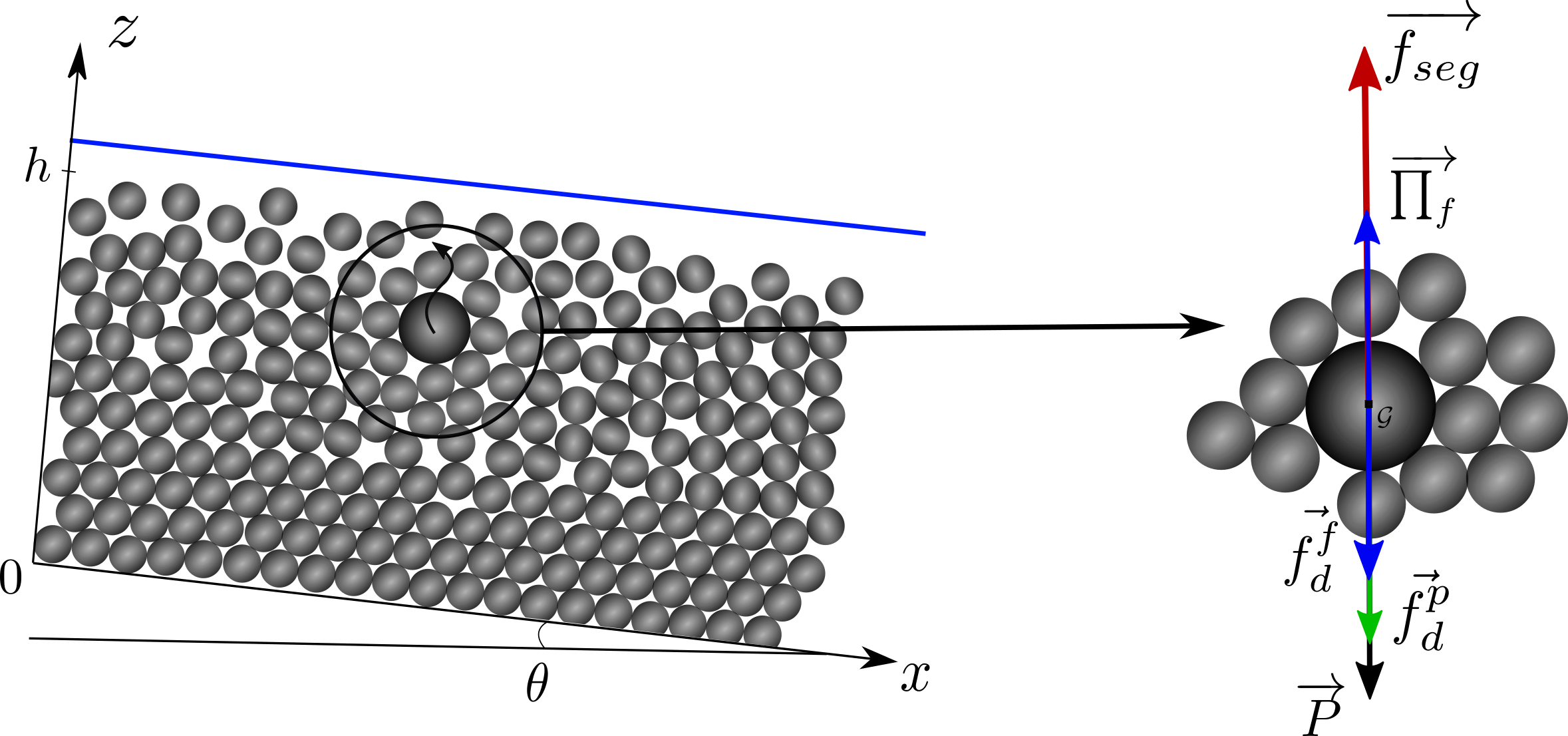}}
	\caption{Vertical component of the forces acting on a large intruder. $\vec{\Pi_f}$ (\textcolor{F3}{\protect\tikz[baseline]{\protect\draw[line width=0.35mm] (0,.5ex)--++(0.65,0) ;}}) is the buoyancy due to the fluid and $\vec{f_{seg}}$ (\textcolor{F1}{\protect\tikz[baseline]{\protect\draw[line width=0.35mm] (0,.5ex)--++(0.65,0) ;}}) is the segregation force identified by \citet{GuillardScalinglawssegregation2016}. The particle is also submitted to the drag forces $\vec{f^p_{d}}$ (\textcolor{F2}{\protect\tikz[baseline]{\protect\draw[line width=0.35mm] (0,.5ex)--++(0.65,0) ;}}) and  $\vec{f^f_{d}}$ (\textcolor{F3}{\protect\tikz[baseline]{\protect\draw[line width=0.35mm] (0,.5ex)--++(0.65,0) ;}}) respectively due to the interaction with small particles \citep{TripathiDensitydifferencedrivensegregation2013} and the fluid.}
	\label{single_in_smalls}
\end{figure}
The configuration is sketched in figure \ref{single_in_smalls}. The large particle is of diameter $d_l$, of volume $V_l$ and of density $\rho^p$ in a bed of height $h$ made of small particles. Below this layer the grains are in the quasi-static regime. 
Applying the Newton's second law, the vertical Lagrangian equation of the intruder can be expressed as
\begin{equation}
\rho^p V_l\dfrac{d w^l}{dt} = P - \mathbf{\Pi}_f + f^f_{d} + f^p_{d} - f_{seg}.
\label{lagrangian_eq}
\end{equation} 
In equation \ref{lagrangian_eq}, the large intruder is submitted to five forces (see figure \ref{single_in_smalls}): its weight $P = - \rho^p V_l g \cos \theta$, the buoyancy force $\mathbf{\Pi}_f = -\rho^f V_l g \cos \theta$, the segregation force $f_{seg}$, the drag force exerted by the fluid $f_d^f$ and the drag force exerted by the small particles $f_d^p$. 
In the present configuration, the slope angle is low ($\tan\theta = 0.1$) and the streamwise gravity component is negligible, making the contribution of the shear stress gradient to the segregation force negligible (see equation \ref{segregationForce_intro} from \citet{GuillardScalinglawssegregation2016}). Thus, the pressure gradient contribution is dominant and the segregation force can be simplified as 
\begin{equation}
f_{seg} = V_l \mathcal{F}(\mu) \dd{P^s}{z},
\label{segregationForce}
\end{equation}
where the function form $\mathcal{F}(\mu)$ will be called the empirical segregation function.\\

While segregating at a velocity $w^l$, the large particle is submitted to a fluid drag force. Because the particulate Reynolds number based on the vertical velocity $Re_p = d_l \rho^f w^l/ \eta^f$ is very small in the bed, fluid inertial effects are negligible at the particle length scale and the vertical fluid drag force $f^{d}_f$ may be approximated by the Stokes law \citep{StokesMathematicalPhysicalPapers1851}
\begin{equation}
f^f_{d} = 3 \pi \eta^f d_l (w^f - w^l),
\label{single_fluid_drag_force}
\end{equation}
where $\eta^f$ is the fluid dynamics viscosity and $w^f$ is the vertical velocity of the fluid. \\

During its segregation motion, the intruder is also submitted to frictional forces from the surrounding small particles. This results in a particle drag force $f^p_{d}$ modeled as proposed by \citet{TripathiDensitydifferencedrivensegregation2013} (see equation \eqref{solid_drag_intro}) as 
\begin{equation}
f^{p}_{d} = c \pi \eta^{p} d_l \big(w^{s}-w^l\big),
\label{solidDrag}
\end{equation}
where $w^s$ is the vertical velocity of small particles and the drag coefficient $c$ is first approximated as a constant equal to 3 \citep{TripathiDensitydifferencedrivensegregation2013}.

\subsection{Dimensionless equation for the large intruder}
In order to identify the dominant terms in equation \eqref{lagrangian_eq}, it is made dimensionless using classical scalings for granular flows:
\begin{equation}
w^k = \sqrt{d_l g} \tilde{w}^k,  \quad z = d_l  \tilde{z},  \quad t = \sqrt{d_l / g} \tilde{t} \quad \text{and} \quad p^k = \rho^p d_l g \tilde{p^k},
\label{dimensionless_variables}
\end{equation}
where $k=s, l \text{ or } f$ respectively for the surrounding small particles, the large intruder and the fluid.
Introducing these variables in \eqref{lagrangian_eq} and taking into account that $\cos \theta \sim 1$, the dimensionless form of the large intruder Lagrangian equation of motion can be written as
\begin{equation}
\dfrac{d \tilde{w^l}}{d\tilde{t}} = -\dfrac{\rho^p-\rho^f}{\rho^p} + \dfrac{\tilde{w^f}-\tilde{w^l}}{St^f} + \dfrac{\tilde{w^s}-\tilde{w^l}}{St^p} -  \mathcal{F}(\mu) \dd{\tilde{P^s}}{\tilde{z}}.
\label{dimensionless_lagrangian_eq}
\end{equation}
Equation \eqref{dimensionless_lagrangian_eq} contains two dimensionless numbers. The first one is the fluid Stokes number
\begin{equation}
St^f = \dfrac{\rho^p d_l W}{6 c \eta^f},
\label{fluid_stokes_number}
\end{equation}
in which $W=\sqrt{d_l g}$ is the characteristic velocity of the large particle. This Stokes number compares the inertia of the large intruder with the viscous friction exerted by the fluid. Similarly, the granular Stokes number
\begin{equation}
St^p = \dfrac{\rho^p d_l W}{6 c \eta^p},
\label{particle_stokes_number}
\end{equation}
compares the inertia of the large intruder with the contact friction exerted by the small particles in the vicinity of the intruder.\\

Assuming a classical bedload configuration, the fluid flows inside the porous matrix of the granular bed. Only the first layer of particles at the top is in a dense flow regime. Below this layer the grains are in the quasi-static regime. 
Typical values of the granular viscosity for dense granular flows are very high compared with the water viscosity (typical ranges span from $10^3$ at the bed surface to $10^6$ at the bed bottom). This results in a fluid Stokes number $St^f$ much larger than the granular Stokes number $St^p$ whatever the height into the bed. Therefore, the second term in the right hand side of equation \eqref{dimensionless_lagrangian_eq}, representing the fluid drag force, can be neglected. In addition, while the large intruder is rising, it only modifies the small particle bed structure locally. Therefore, it is assumed that the averaged vertical velocity of the small particles $w^s$ is negligible compared with $w^l$.
Focusing on the position of the intruder in the quasi-static part of the bed, it can be deduced from equation \eqref{dimensionless_lagrangian_eq} that the total solid volume fraction is constant, $\Phi = cste$. Therefore, for this configuration, a simple equation for the vertical velocity of the large intruder can be written as
\begin{equation}
\dfrac{d \tilde{w^l}}{d\tilde{t}} + \dfrac{1}{St^p} \tilde {w^l} = \dfrac{\rho^p - \rho^f}{\rho^p} \left(\Phi \mathcal{F}(\mu) - 1 \right).
\label{lagrangian_neglected_terms}
\end{equation}
This dimensionless equation does not depend on the fluid parameters and should be valid to model the vertical velocity of an intruder segregating in a dry granular flow.
In addition, with this equation, it can be shown from an asymptotical analysis that the formulation \eqref{function_Guillard} of the empirical segregation function $\mathcal{F(\mu)}$ does not satisfy equation \eqref{dimensionless_lagrangian_eq} when small particles are not moving (see appendix \ref{AppendixA} for more details). In order to satisfy this constraint, it is proposed to use the following functional form
\begin{equation}
\mathcal F(\mu) = \dfrac{1}{\Phi_{max}} + \big(1-e^{-70(\mu-\mu_c)}\big),
\label{revisited_Fmu}
\end{equation}
which has the same range of values than the original one but satisfies equation \eqref{dimensionless_lagrangian_eq} for no flow condition.
Equation \eqref{lagrangian_neglected_terms} allows one to identify the main size segregation mechanisms and shows that the segregation of a large intruder can be seen as a simple relaxation process with characteristic time $St^p$.
	
\section{Volume averaged multi-phase flow model}
\label{multi_phase_section}
As discussed in the previous section, the dynamics of a large intruder in an immersed granular flow made of small particles can be described using interparticle forces published in the literature. In this section, the goal is to upscale this result by volume-averaging the segregation forces over a collection of large particles in order to make the link between this discrete picture and continuum models for size segregation. This is done in the framework of the volume averaged equations from \citet{JacksonLocallyaveragedequations1997, Jacksondynamicsfluidizedparticles2000} which provides continuum equations for the three phases: large particles, small particles and the interstitial fluid.
\subsection{3D general governing equations}
\label{GeneralModelling}
Following \citet{JacksonLocallyaveragedequations1997, Jacksondynamicsfluidizedparticles2000}, the mass and momentum balance equations for each class are given by

\begin{equation}
\dd{\epsilon \rho^f \textbf{u}^f}{t} + \mathbf{\nabla} . \left(\epsilon \rho^f \textbf{u}^f\right),
\label{GlobalMassFluid}
\end{equation}
\begin{equation}
\dd{\Phi^i \rho^p  \textbf{u}^i}{t} + \mathbf{\nabla} . \left( \Phi^i \rho^p \textbf{u}^i\right),
\end{equation}

\begin{equation}
\dd{\epsilon \rho^f  \textbf{u}^f}{t} + \nabla . \left( \epsilon \rho^f \textbf{u}^f \otimes \textbf{u}^f \right)  = \mathbf{\nabla}.\textbf{S}^f - \epsilon \rho^f  \textbf{g} - n_l \textbf{f}_{f\rightarrow l} - n_s\textbf{f}_{f\rightarrow s} 
\label{GlobalMomFluid}
\end{equation}
\begin{equation}
\dd{\Phi^i \rho^p  \textbf{u}^i}{t} + \nabla . \left( \Phi^i \rho^p  \textbf{u}^i \otimes \textbf{u}^i \right) = \mathbf{\nabla}.\textbf{S}^i - \Phi^i \rho^p  \textbf{g} + n_i \textbf{f}_{f\rightarrow i}  + n_i \textbf{f}_{\delta\rightarrow i},
\label{GlobalMomParticles}
\end{equation}
where $f$ is the fluid and indices $i=l,s$ denote the large particle phase and the small particle phase respectively ($\delta = l$ if $i = s$ and  $\delta = s$ if $i = l$). $\Phi^l$ and $\Phi^s$ are the volume fractions for the large and small grains and verify $\Phi^s + \Phi^l = \Phi$ where $\Phi$ is the volume fraction of the mixture, i.e. the total solid volume fraction. Consequently, the fluid volume fraction is $\epsilon = 1- \Phi^l - \Phi^s$.
$\mathbf{S}^k$ is the stress tensor associated with phase $k$ with $k = l, s \text{ or } f$. They can be separated into pressure and shear stress contribution
\begin{equation}
\mathbf{S}^k = -p^k \mathbf{I} + \mathbf{\tau}^k,
\end{equation}
where $\mathbf{\tau}^k$ is the shear stress tensor and $p^k$ is the pressure of phase $k$. It should be noticed that for a solid phase, the static pressure arises from the enduring contacts between the particles. Thanks to the mixture model approaches \citep{MorlandFlowviscousfluids1992}, it can be assumed that each particle phase carries the total overburden pressure $p^m$ according to their local volume fraction as
\begin{equation}
p^i = \dfrac{\Phi^i}{\Phi} p^m,
\label{partialPressure}
\end{equation}
where $m$ denotes the mixture made of small and large particles. The total overburden pressure $p^m$is computed using the formulation proposed by \citet{JohnsonFrictionalcollisionalconstitutive1987}. For further information, the reader is referred to \citet{ChauchatSedFoam23Dtwophase2017} and \citet{Chauchatcomprehensivetwophaseflow2018}.\\

The momentum equations \eqref{GlobalMomFluid} and \eqref{GlobalMomParticles} contain two terms coming from the momentum exchange between the different phases: $n_i \textbf{f}_{f\rightarrow i}$ and $n_i \textbf{f}_{\delta\rightarrow i}$. The term $n_i \textbf{f}_{f\rightarrow i}$ is the averaged value of the resultant forces exerted by the fluid on the particles of phase $i$. \citet{Jacksondynamicsfluidizedparticles2000}  showed that for a collection of immersed particles, this interaction force can be written as
\begin{equation}
n_i \textbf{f}_{f\rightarrow i} = - \Phi^i \mathbf{\nabla} p^f + n_i \textbf{f}^{f\rightarrow i}_{d},
\end{equation}
where $\Phi^i \mathbf{\nabla} p^f$ is the buoyancy force exerted by the fluid phase on the particles and $n_i \textbf{f}^{f\rightarrow i}_{d}$ is the particle averaged viscous drag force between the particles and the fluid phase. The term $n_i \textbf{f}_{\delta\rightarrow i}$ is the averaged value of all interacting forces between large and small particle phases. It can be directly expressed in 3D from the local segregation force of \citet{TripathiDensitydifferencedrivensegregation2013} and \citet{GuillardScalinglawssegregation2016}.

Therefore, the developed model is general and can be applied to 3D configurations.  For simplicity and for the purpose of the present study, the model will be only developed for a 1D uniform flow.

\subsection{Simplified 1D vertical multi-phase flow model}
\label{MultiPhaseFlowModel}

The multi-phase flow model (equations \ref{GlobalMassFluid} to \ref{GlobalMomParticles}) is simplified by considering a uniform flow in the streamwise direction. From now, all the variables only depend on the vertical position $z$. Therefore, the spatially averaged velocity of the phase $k$ can be written as $\textbf{u}^k = u^k(z) \textbf{e}_x + w^k(z) \textbf{e}_z$. The mass conservation equations simplify to

\begin{equation}
\dd{\epsilon}{t} + \dd{\epsilon w^f}{z} = 0  \quad \quad \text{and} \quad \quad \dfrac{\partial {\Phi^i}}{\partial {t}} + \dfrac{\partial{{\Phi^i}w^i}}{\partial{z}}=0, 
\label{eq:MassConservation}
\end{equation}
and the momentum balance equations in the vertical direction are:
\begin{equation}
\rho^f \left[\dd{\epsilon w^f}{t}  +  \dd{\epsilon w^f w^f }{z}\right] = -\epsilon \dd{p^f}{z} - \epsilon \rho^f g \cos \theta - n_l<f^{f\rightarrow l}_{d}> - n_s<f^{f\rightarrow s}_{d}>,
\label{MomFluidZ}
\end{equation}
\begin{equation}
\rho^p \left[\dd{\Phi^l w^l}{t}  +  \dd{\Phi^l w^l w^l}{z}\right] = -\dd{p^l}{z} - \Phi^l \dd{p^f}{z} - \Phi^l \rho^p g \cos \theta + n_l<f^{f\rightarrow l}_{d}> + n_l <f_{s\rightarrow l}> ,
\label{MomLargeZ}
\end{equation}
\begin{equation}
\rho^p \left[\dd{\Phi^s w^s}{t}  +  \dd{\Phi^s w^s w^s}{z}\right] = -\dd{p^s}{z} - \Phi^s \dd{p^f}{z} - \Phi^s \rho^p g \cos \theta + n_s<f^{f\rightarrow s}_{d}> + n_s <f_{l\rightarrow s}>.
\label{MomSmallZ}
\end{equation}
In the two last equations, the solid pressures $p^l$ and $p^s$ are given by equation \eqref{partialPressure}.
To solve these equations, it is necessary to prescribe closures for the spatially averaged fluid/grain interaction and grain-grain interactions, and for the granular and fluid pressures.

Considering the fluid-grain interaction, both small and large granular phases interact with the fluid phase through $\Phi^i \nabla p^f$  and the drag force $n_i \textbf{f}^{f\rightarrow i}_{d}$. For an assembly of particles, the spatial averaging of the vertical total drag force applied by the fluid gives
\begin{equation}
n_l <f^{f\rightarrow i}_{d}> = \dfrac{\Phi^i \rho^p }{t_i}\left(w^f-w^i\right),
\label{continuousFluidLargeDrag}
\end{equation}
where  $t_i = \rho^p  d_i^2 (1-\Phi)^{3} /  18 \eta^f$ is the particle response time and $d_i$ is the particle diameter of phase $i$. The factor $(1-\Phi)^{3}$ is a correction proposed by \citet{Richardsonsedimentationsuspensionuniform1954} to take into account hindrance effects.
Since the drag is linear, the spatial averaging is simply the drag force applied on one particle (given in \eqref{single_fluid_drag_force}) multiplied by the number of particles per unit volume   $n_i = \Phi^i/ V^i$ \citep{Jacksondynamicsfluidizedparticles2000}, where $V^i$ is the volume of a single large particle.\\

The granular phases also interact with each other and the grain-grain interaction closure should be prescribed in the model. For a single large grain in a bath of small particles, it has been shown in section \ref{largeIntruderInSmalls} that small particles exert two forces on a large intruder,
\begin{equation}
f_{s\rightarrow l} = f_{d}^p + f_{seg}.	 
\end{equation}
To extend these forces to a collection of large particles, the interaction force $f_{s\rightarrow l}$ is spatially averaged. Since this force is linear, it amounts to multiplying $f_{s\rightarrow l}$ by the number of large particles per unit volume $n_l = 6 \Phi^l / \pi d_l^3$.
Therefore, the total solid interaction force exerted by the small particles on the large ones is given by
\begin{equation}
n_l <f_{s\rightarrow l}> = \dfrac{\Phi^l \rho^p}{t_{ls}} \big(w^s-w^l\big) + \Phi^l \mathcal{F}(\mu) \dd{p^m}{z},
\label{continuousSmallLargeDrag}	 
\end{equation}
where $t_{ls} = \rho^p d_l^2 / 6 c \eta^p$ is the particle response time for the drag force between small and large particles and $\mathcal{F}(\mu) = \big(1-e^{-70(\mu-\mu_c)}\big)$ (see appendix \ref{AppendixB}). According to the Newton's third law, the force exerted by the large particles on the small ones (equation \ref{MomSmallZ}) is
\begin{equation}
n_s <f_{l\rightarrow s}> = - n_l <f_{s\rightarrow l}>.
\label{3_Newton_law}
\end{equation}\\
The solid mixture phase is made of both particle phases and is noted with $i=m$. Its momentum balance is obtained by summing \eqref{MomLargeZ} and \eqref{MomSmallZ}. Since the mixture phase does not distinguish between small and large particles, the solid interaction forces should not appear in this equation. Equation \eqref{3_Newton_law} ensures that these forces vanish when developing the mixture momentum equation.\\

The proposed volume-averaged multi-phase flow model describes size segregation of a bi-disperse mixture immersed in a fluid. This represents an improvement upon the model of \citet{Thorntonthreephasemixturetheory2006}, which was based on semi-empirical parametrisation of the interparticle forces between small and large particles. The present model provides closures based on forces applied on a single particle, and bridges the gap between granular scale processes and continuum modelling in size segregation.
This important result will be used in the following to derive an advection-diffusion model for size segregation.  

\section{Derivation of the advection-diffusion model}
\label{section_adv_diff}
A classical continuum approach to model size segregation is the advection-diffusion model \citep{DolguninSegregationmodelingparticle1995,Graytheoryparticlesize2005}. These models can be derived from the multicomponent mixture theory \citep{Thorntonthreephasemixturetheory2006, gray_ancey_2011} by substituting the percolation velocity of one particle-size into the mass conservation equation. The advection and diffusion coefficients can be modelled using experimental and theoretical closures \citep{DolguninDevelopmentmodelsegregation1998,vanderVaartUnderlyingAsymmetryParticle2015, FerdowsiRiverbedarmouringgranular2017,CaiDiffusionsizebidisperse2019} or can be derived as a simplification from the continuum model of \citet{Thorntonthreephasemixturetheory2006} and \citet{GrayParticlesizesegregationdiffusive2006}. In the present section, the multi-phase model developed in the previous section (equations \ref{MomFluidZ} to \ref{MomSmallZ}) makes it possible to derive an advection-diffusion model similar to \citet{Thorntonthreephasemixturetheory2006} and \citet{GrayParticlesizesegregationdiffusive2006}, with advection and diffusion coefficients depending on the segregation and the drag forces \citep{GuillardScalinglawssegregation2016, TripathiDensitydifferencedrivensegregation2013} determined in independent idealised configurations.\\

Combining equations \eqref{MomSmallZ}, \eqref{continuousFluidLargeDrag}, \eqref{continuousSmallLargeDrag} and \eqref{3_Newton_law}, the momentum balance of small particles can be written as:
\begin{multline}
\rho^p \left[\dd{\Phi^s w^s}{t}  +  \dd{\Phi^s w^s w^s}{z}\right] = -\dd{p^s}{z} - \Phi^s \dd{p^f}{z} - \Phi^s \rho^p g \cos \theta + \dfrac{\rho^p \Phi^s}{t_s}\left(w^f-w^s\right) \\- \dfrac{\rho^p \Phi}{t_{ls}} \big(w^s-w^m\big) + \Phi^l \mathcal{F}(\mu) \dd{p^m}{z}.
\label{MomSmallZ_2}
\end{multline}	
The total volume fraction $\Phi = \Phi^s + \Phi^l$ is assumed to be constant and equal to $\Phi_{max} = 0.61$ since particle velocity fluctuations are small.
For a deposited bed, the particle momentum balance in the wall-normal direction reduces to a hydrostatic pressure distribution for both the fluid and the particle phases \citep{Chauchatcomprehensivetwophaseflow2018}:
\begin{equation}
\dd{p^f}{z} = -\rho^f g\cos \theta \quad \text{and} \quad \dd{p^m}{z} = -\Phi \left(\rho^p - \rho^f\right) g \cos \theta.
\label{hydrostatic_pressure_gradient}
\end{equation}
Assuming a constant mixture solid phase volume fraction, the pressure gradient can be integrated to give the pressure distributions:
\begin{equation}
p^f = \rho^f g\cos \theta \left(h-z\right) \quad \text{and} \quad p^m = \Phi \left(\rho^p - \rho^f\right) g \cos \theta \left(h-z\right).
\label{hydrostatic_pressure}
\end{equation}
Following \citet{Thorntonthreephasemixturetheory2006}, the volume fraction per unit granular volume is introduced as $\phi^i = \Phi^i/\Phi$. This notation is more convenient since it ensures $\phi^s + \phi^l = 1$.
Using equation \eqref{partialPressure}, the momentum equation \eqref{MomSmallZ_2} for small particles is rewritten as follows:
\begin{multline}
\Phi \rho^p \left[\dd{\phi^s w^s}{t}  +  \dd{\phi^s w^s w^s}{z}\right] = -p^m \dd{\phi^s}{z} + \dfrac{\rho^p \phi^s \Phi }{t_s}\left(w^f-w^s\right) \\ - \dfrac{\rho^p \Phi}{t_{ls}} \big(w^s-w^m\big) + \phi^l \Phi \mathcal{F}(\mu) \dd{p^m}{z}.
\label{MomSmallZ_3}
\end{multline}
Using the same scalings as in the Lagrangian equation \eqref{dimensionless_lagrangian_eq} for a single intruder, the equation \eqref{MomSmallZ_3} is made dimensionless as follows:
\begin{equation}
\dfrac{\partial \phi^s\tilde{w}^{s}}{\partial {\tilde{t}}} +\dd{ \phi^s \tilde{w}^{s}\tilde{w}^{s}}{\tilde{z}}= - \dfrac{\tilde{p}^m}{\Phi} \dd{\phi^s}{\tilde{z}}  + \dfrac{\phi^s}{St^f} \left(\tilde{w}^f -\tilde{w}^{s}\right)  - \dfrac{\left(\tilde{w}^s -\tilde{w}^{m}\right)}{St^p} +\phi^l \mathcal{F}(\mu) \dd{\tilde{p}^m}{\tilde{z}}.
\label{dimensionlessEquation}
\end{equation}
As shown in section \ref{largeIntruderInSmalls}, $St^f>>St^p$ in the bed and the fluid drag force can be neglected. Furthermore, assuming a quasi-steady state and neglecting inertial terms, equation \eqref{dimensionlessEquation} can be rewritten as
\begin{equation}
 - \dfrac{\tilde{p}^m}{\Phi} \dd{\phi^s}{\tilde{z}}  - \dfrac{\left(\tilde{w}^s -\tilde{w}^{m}\right)}{St^p} +\phi^l \mathcal{F}(\mu) \dd{\tilde{p}^m}{\tilde{z}} = 0,
\label{dimensionlessEquation2}
\end{equation}
from which, assuming $w^m =0$ \citep{Thorntonthreephasemixturetheory2006}, the flux of small particles can be expressed as
\begin{equation}
\phi^s \tilde{w}^s = - \dfrac{\phi^s}{\Phi} \tilde{p}^m St^p \dd{\phi^s}{\tilde{z}} + \phi^l \phi^s \mathcal{F}(\mu) St^p  \dd{\tilde{p}^m}{\tilde{z}}.
\label{small_flux}
\end{equation}
Equation \eqref{small_flux} is then substituted in the mass conservation equation \eqref{eq:MassConservation} to obtain the following advection-diffusion equation for the percolation of small particles:
\begin{equation}
\dfrac{\partial \phi^s}{\partial \tilde{t}} + \dfrac{\partial}{\partial \tilde{z}}\big(\phi^l \phi^s S_r \big) =   \dfrac{\partial}{\partial \tilde{z}}\big(D \dfrac{\partial \phi^s}{\partial \tilde{z}}  \big),
\label{adv-diff}
\end{equation}
with 
\begin{equation}
S_r = \mathcal{F(\mu)} St^p \dd{\tilde{p}^m}{\tilde{z}} \quad \text{and} \quad D = \dfrac{ \phi^s \tilde{p}^m St^p}{\Phi},
\label{SegregationNumber_DiffusionCoefficient}
\end{equation}
Since the pressure gradient is negative, $S_r$ is negative which ensures a downward flux for the small particle phase.

In equation \eqref{adv-diff}, $S_r$ is the segregation number or advection coefficient and $D$ is the diffusion coefficient. Here, equation \eqref{SegregationNumber_DiffusionCoefficient} provides physical closures, which are directly obtained  from volume-averaging of the particle scale segregation forces. The advection coefficient $S_r$ is therefore expressed as a product between the segregation term $\mathcal{F(\mu)} \partial \tilde{p}^m / \partial \tilde{z}$, which quantifies the ability for the small particles to fall downward, and the solid Stokes number which quantifies the drag force exerted by the other grains counteracting this downward movement.
It is interesting to note that the granular Stokes number is present in both coefficients which indicates that it is a key parameter for the advection and the diffusive remixing. Equation \eqref{SegregationNumber_DiffusionCoefficient} shows that the mixture pressure is also an important parameter for diffusion.

These results improve upon the original model from \citet{Thorntonthreephasemixturetheory2006} and \citet{GrayParticlesizesegregationdiffusive2006} since the experimentally-based closures \citep{GuillardScalinglawssegregation2016, TripathiDensitydifferencedrivensegregation2013} make it possible to link both the advection and the diffusion coefficients to the local physical parameters of the granular flow. This result not only provides closures for the advection-diffusion model based on local granular forces but also highlights the key local physical mechanisms controlling segregation and diffusion. Since the advection and diffusion coefficients do not contain fluid parameters, it indicates that even though the equation has been set for a bedload configuration, it is also valid for dry granular flows. 

In the following, the relevance of the obtained advection-diffusion model with respect to the multi-phase flow model will be tested. 
\section{Comparison with existing discrete numerical simulations}
\label{results}
In this section, the multi-phase flow model presented in section \ref{multi_phase_section} and the corresponding advection-diffusion model presented in section \ref{section_adv_diff} is tested against the discrete numerical simulations of \citet{Chassagne2020}.
This study focused on the segregation of small particles initially resting on top of large ones. It provides a comprehensive dataset to evaluate local granular parameters such as the volume fraction and the segregation velocities. In addition, since it is not the purpose of this work to develop a granular rheology, the shear stress and shear rate profiles obtained from the DEM will be used as input parameters for the continuum models.\\
The 3D DEM configuration and the main results from \citet{Chassagne2020} are first summarised (section \ref{Summary_DEM_part}). Then, in section \ref{comparison_DEM_part}, the multi-phase flow model is compared with the DEM results using default parameters (see sections \ref{multi_phase_section} and \ref{section_adv_diff}) for the segregation of the small particles. Finally, in section \ref{compare_models}, the results predicted by the advection-diffusion model and the multi-phase flow model are compared to determine the validity of the former with respect to the latter.
\subsection{DEM investigation of \citet{Chassagne2020}}
\label{Summary_DEM_part}
\begin{figure}
	\hspace{-1.5cm}
	\begin{subfigure}[t]{0.362\textwidth}
		\centering \includegraphics[width=\textwidth]{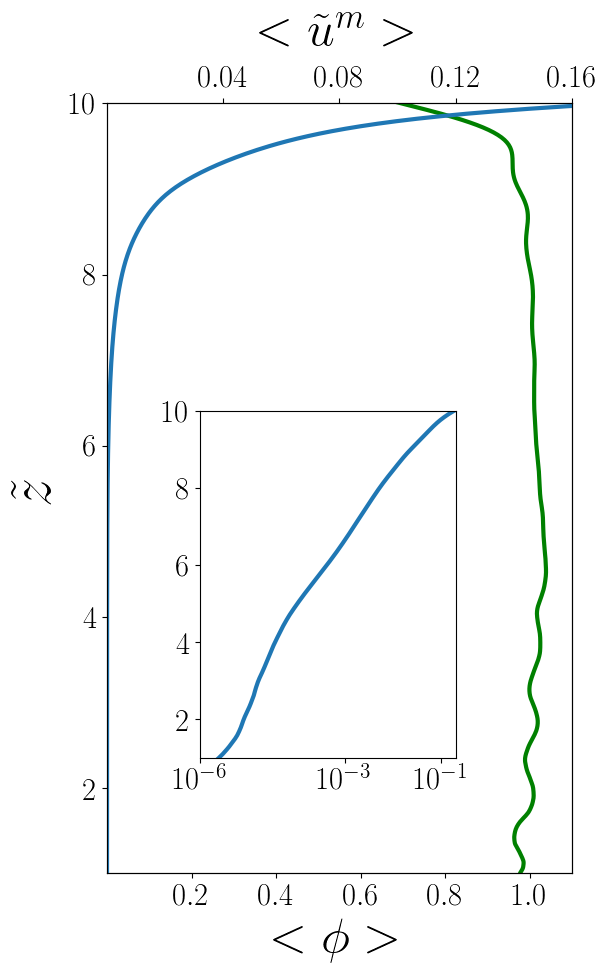}
		\caption{}\label{velocity_c}
	\end{subfigure}
	\begin{subfigure}[t]{0.32\textwidth}
		\centering \includegraphics[width=\textwidth]{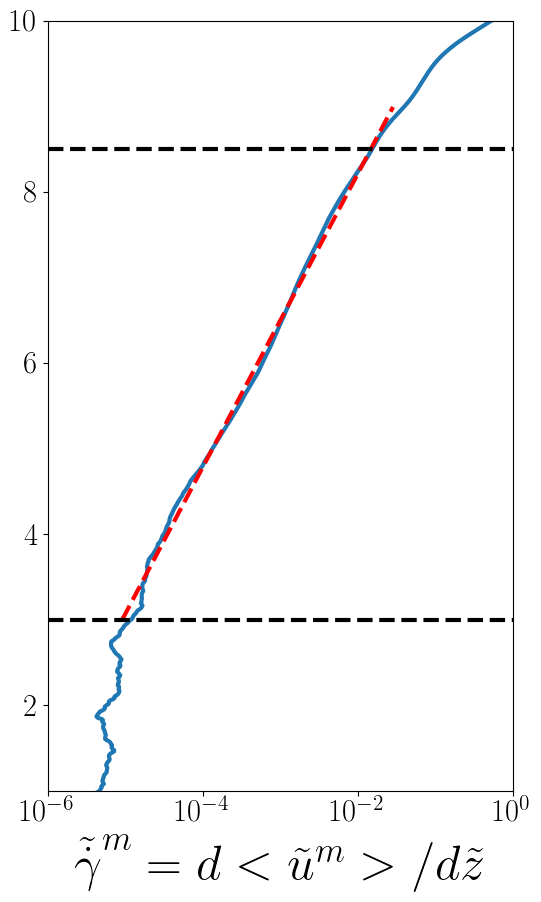}
		\caption{}\label{fit_gamma_a}
	\end{subfigure}
	\begin{subfigure}[t]{0.32\textwidth}
		\centering \includegraphics[width=\textwidth]{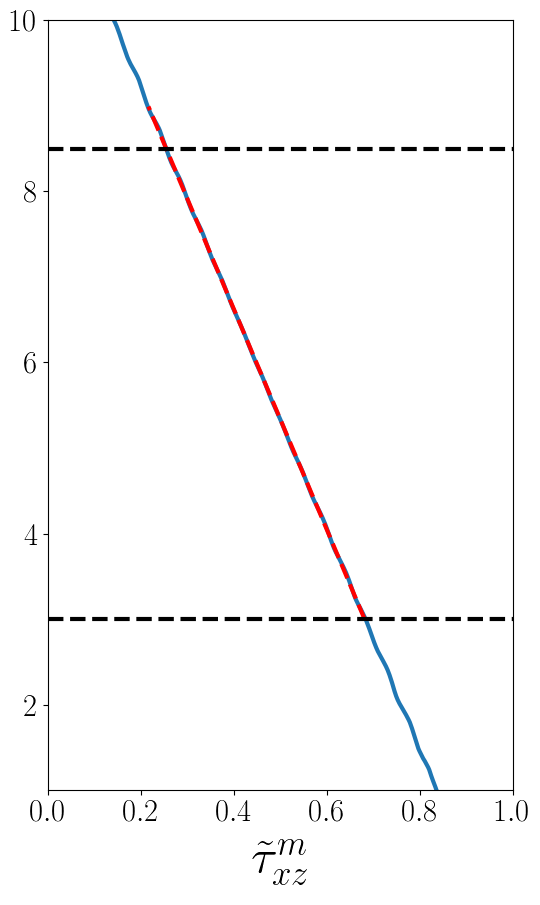}
		\caption{}\label{fit_tau_b}
	\end{subfigure}
	\begin{subfigure}[b]{0.55\textwidth}
		\includegraphics[width=\textwidth]{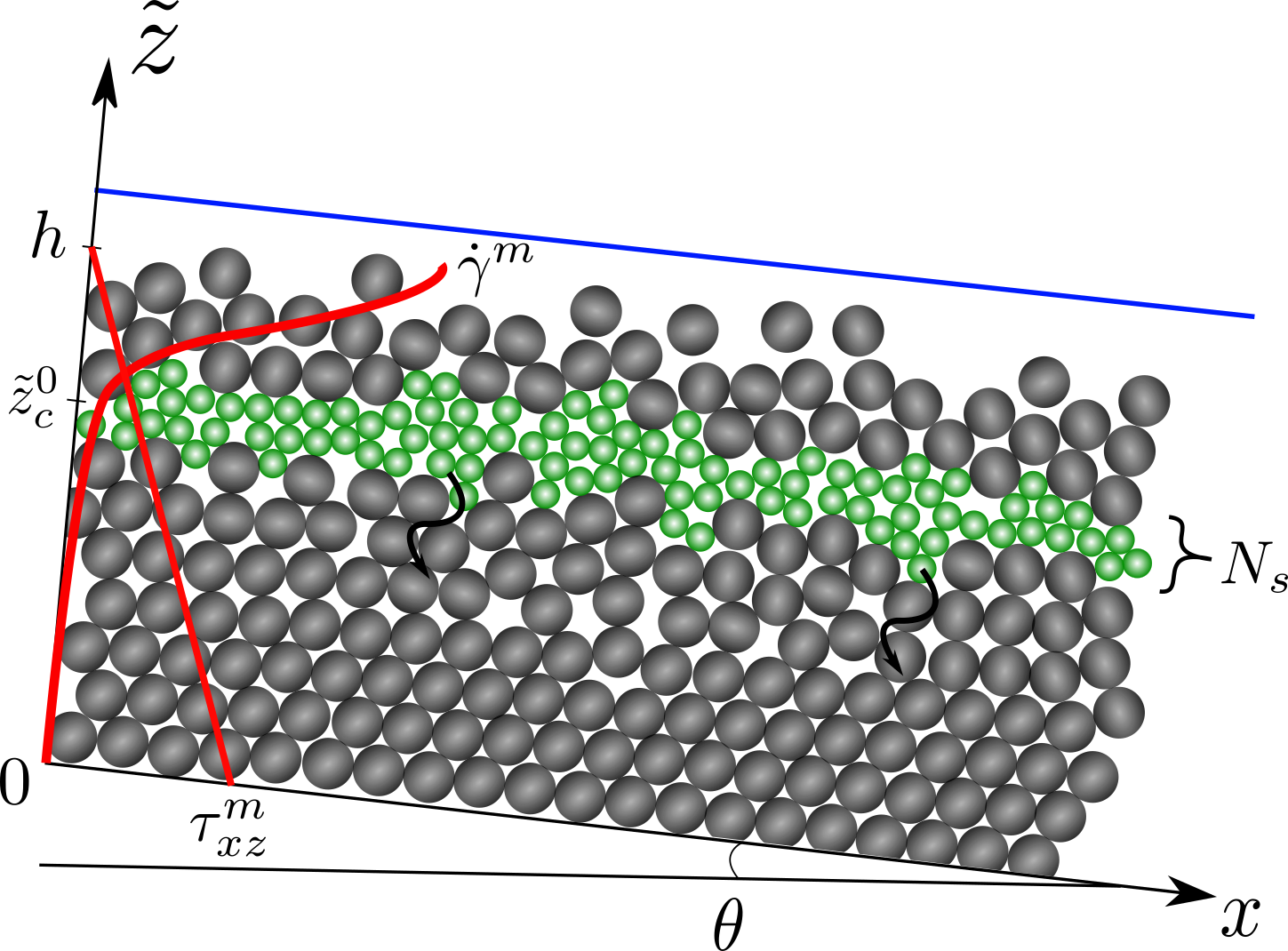}
		\caption{}\label{numerical_set_up}
	\end{subfigure}
	\begin{subfigure}[b]{0.45\textwidth}
		\includegraphics[width=\textwidth]{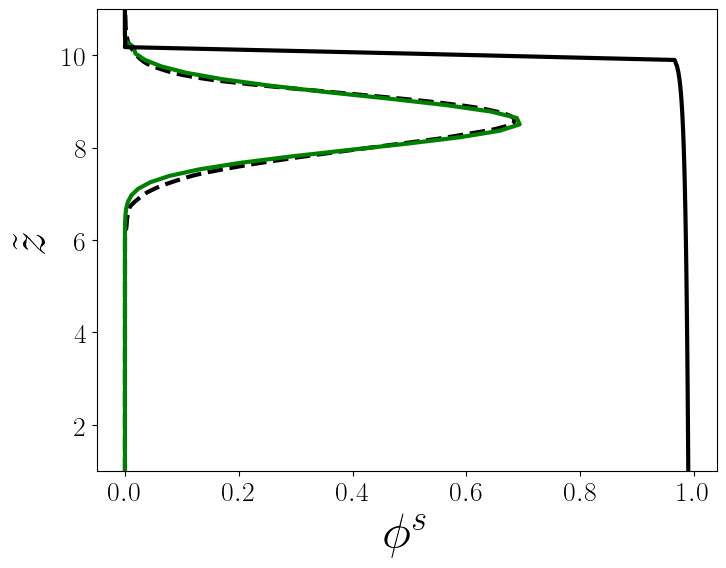}
		\caption{}\label{concentration_init}
	\end{subfigure}
	\caption{\label{fit_gamma} Profiles and configuration from the DEM simulations. (a) Streamwise mixture velocity profile in the bed (\textcolor{C2}{\protect\tikz[baseline]{\protect\draw[line width=0.35mm] (0,.5ex)--++(0.65,0) ;}}) and mixture volume fraction (\textcolor{C1}{\protect\tikz[baseline]{\protect\draw[line width=0.35mm] (0,.5ex)--++(0.65,0) ;}}). The inset is the semilog plot of the velocity profile. (b) Solid shear rate (\textcolor{C2}{\protect\tikz[baseline]{\protect\draw[line width=0.35mm] (0,.5ex)--++(0.65,0) ;}}) and the corresponding fit $\tilde{\dot{\gamma}}^m = \gamma_0 e^{\tilde{z}/s_0}$ (\textcolor{C0}{\protect\tikz[baseline]{\protect\draw[line width=0.35mm,densely dashed] (0,.5ex)--++(0.65,0) ;}}) with $\gamma_0 = 1.64 \times 10^{-7}  $ and $s_0 = 0.74$. (c) Solid shear stress and the corresponding fit $\tilde{\tau}_{xz}^m = a_0 \tilde{z} + \tau_0$ (\textcolor{C0}{\protect\tikz[baseline]{\protect\draw[line width=0.35mm,densely dashed] (0,.5ex)--++(0.65,0) ;}}) with $a_0=-0.078$ and $\tau_0=0.91$.  The top and lower boundary of the quasi-static bed are represented by {\protect\tikz[baseline]{\protect\draw[line width=0.35mm,densely dashed] (0,.5ex)--++(0.65,0) ;}}. (d) Sketch of the numerical experiment with the input profiles for the rheology. (e) Concentration profile of small particles at the initial state for the multi-phase flow model (\textcolor{C1}{\protect\tikz[baseline]{\protect\draw[line width=0.35mm] (0,.5ex)--++(0.65,0) ;}}), the DEM ({\protect\tikz[baseline]{\protect\draw[line width=0.35mm,densely dashed] (0,.5ex)--++(0.65,0) ;}}) and the mixture concentration profile $\phi$ ({\protect\tikz[baseline]{\protect\draw[line width=0.35mm] (0,.5ex)--++(0.65,0) ;}}).}
\end{figure}

In this section, the configuration explored and the main results of \citet{Chassagne2020} are briefly presented. The authors investigated grain-size segregation in turbulent bedload transport using a coupled fluid-Discrete Element Model (DEM) originally developed by \citet{Maurinminimalcoupledfluiddiscrete2015}. For further details, the interested reader is referred to \citet{Chassagne2020}.
The 3D bi-periodic DEM set-up consisted in depositing a layer of small particles over large ones, and letting the particles entrained by the fluid flow at a fixed Shields number. The latter is the dimensionless fluid bed shear stress $Sh = \tau^f /[(\rho^p-\rho^f)g d_l]$ and was taken equal to $0.1$.  The bed slope was fixed to $10\%$, which is representative of mountain streams.
The size ratio was taken as $r=1.5$ with small particles of diameter $d_s = 4mm$ and large particles of diameter $d_l = 6mm$. 
The amounts of large and small particles were assimilated to a number of layers, $N_l$ and $N_s$. The number of layers of a given class represents the height, in terms of particle diameter of this class, occupied by particles if the concentration was equal to the random close packing ($\Phi_{max} = 0.61$). In this way, the bed height at rest was defined as $h = N_l d_l + N_s d_s$ and was fixed to $h=10 d_l$ with a random close packing volume fraction $\Phi_{max} = 0.61$  (profile in figure \ref{velocity_c}). In the study of \citet{Chassagne2020}, different simulations have been performed with $N_s$ varying from $0.01$ (a few isolated particles) to $N_s=2$. In this section it was decided that the comparison would be made with $N_s = 1.5$. The bulk response of the granular mixture to this fluid forcing is represented by the dimensionless mixture streamwise velocity profile in figure \ref{velocity_c}. The inset is a semilog plot of the dimensionless velocity profile and shows that it is exponential. As shown in figure \ref{fit_gamma_a} the linearity of the curve in the semilog plot confirms that the shear rate is exponentially decreasing in the quasi-static part of the bed (delimited by the two horizontal black dashed lines). As expected for a uniform flow, the mixture shear stress $\tilde{\tau}_{xz}^m$ shown in figure \ref{fit_tau_b} is linear with depth. For both quantities, the following fits were proposed and plotted as a red dotted line in figure \ref{fit_gamma}.\\
\begin{equation}
\tilde{\dot{\gamma}}^p = \gamma_0 e^{\tilde{z}/s_0} \quad \quad \text{and} \quad \quad \tilde{\tau}^p_{xz} = a_0 \tilde{z} + \tau_0
\label{equation_fit}
\end{equation}
The simulations performed by \citet{Chassagne2020} on this configuration were focused on the downward segregation of small particles. It was observed that the layer of small particles percolates rapidly for $\tilde{z}> 8.5$ (flowing layer) and then slows down below (this limit is marked in figures \ref{fit_gamma_a} and \ref{fit_tau_b}). \citet{Chassagne2020} showed that the small particles are advected downward like a travelling wave into the bed made of large particle with a layer of constant thickness. As figure \ref{concentration_init} shows, the small particle concentration has a Gaussian-like shape and remains self-similar in time while segregating. The center of mass of small particles, $\tilde{z}_c$, is therefore representative of the dynamics of the entire layer. \citet{Chassagne2020} observed that the small particle layer travels down as a logarithmic function of time:
\begin{equation}
\tilde{z}_c(t) = -a_1 \ln \tilde{t} + b,
\label{log_descent_formula}
\end{equation} 
where $a_1$ is a constant characterising the segregation velocity ($d\tilde{z}_c(t)/d \tilde{t} = - a_1/\tilde{t}$). The authors demonstrated that this logarithmic descent of small particles is a consequence of the dependency of the segregation velocity on the inertial number as
\begin{equation}
\dfrac{d \tilde{z}_c}{d t} \propto I^{0.85}(\tilde{z}_c).
\label{velocity_propto_I}
\end{equation}
Using equation \eqref{velocity_propto_I} in the framework of the advection-diffusion model of \citet{Thorntonthreephasemixturetheory2006} and \citet{GrayParticlesizesegregationdiffusive2006} (see equation \ref{adv-diff_intro}) it was shown that the advection coefficient could be written as
\begin{equation}
S_{r} = S_{r0} I^{0.85},
\label{SegregationNumber}
\end{equation}
where $S_{r0} =  0.049$.
Then, with the help of a travelling wave method, they evidenced that the small particles percolate as a layer and with a self-similar concentration profile because the ratio between the advection coefficient and the diffusion coefficient is constant. The P\'eclet number reads 
\begin{equation}
Pe = \dfrac{S_{r}}{D},
\label{Pe_DEM}
\end{equation}
and is therefore constant with $\tilde{z}$, so that the diffusion coefficient has to have the same dependency on the inertial number as the segregation coefficient:
\begin{equation}
D=D_{0} I^{0.85},
\label{DiffusionCoefficient_DEM}
\end{equation}
where $D_0$ is taken as $D_0 = 0.01$.

This work also demonstrated that the dynamics of the fine particle layer is controlled by its bottom position, which acts as a lower bound for the segregation velocity. In this way, the particles in the layer cannot segregate faster.\\

The numerical resolution of the 1D multi phase model $a priori$ requires to solve the granular rheology in order to estimate the granular viscosity required to evaluate the granular drag force contribution. The goal of the present study is to focus on grain-size segregation modelling. In addition, the results  of \citet{Chassagne2020} were obtained in the quasi-static regime, on which there is still no concensus regarding granular rheology. For these reasons, the granular viscosity is directly determined by the DEM results here. This makes it possible to focus on the effect of the segregation model and to put aside potential discrepancies linked to a non-accurate description of the granular rheology. 
The granular viscosity is therefore computed from the DEM results using the definition:
\begin{equation}
\eta_p = \dfrac{\tau^m_{xz}}{\big \vert \dot{\gamma^m_{xz}}\big \vert}.
\label{granularviscosity}
\end{equation}
As shown by the red dashed lines in figures \ref{fit_gamma_a} and \ref{fit_tau_b}, the fits presented in equation \ref{equation_fit} match the DEM results in the region $8.5>\tilde{z}>3$. Thus, using these expressions in eq (\ref{granularviscosity}) provides an accurate estimate of the granular viscosity for the particle-particle drag closure. For this reason, the validation of the multi-phase flow model will only be carried out in this part of the bed. Therefore, the initial state consists in placing the small particles in the upper limit of the quasi-static part with the center of mass $\tilde{z}^0_c = 8.5$. The sketch of this configuration is shown in figure \ref{numerical_set_up}. Figure \ref{concentration_init} shows the small particle concentration profile of this numerical set-up at the initial state. The initial concentration is taken with a gaussian fit on the DEM initial concentration (figure \ref{concentration_init}), and ensures that the mass of particles is the same in the DEM and in the continuum simulations.

\subsection{Comparison with the multi-phase flow model}
\label{comparison_DEM_part}
\begin{figure}	
	\begin{subfigure}[t]{0.48\textwidth}
		\centering \includegraphics[width=\textwidth]{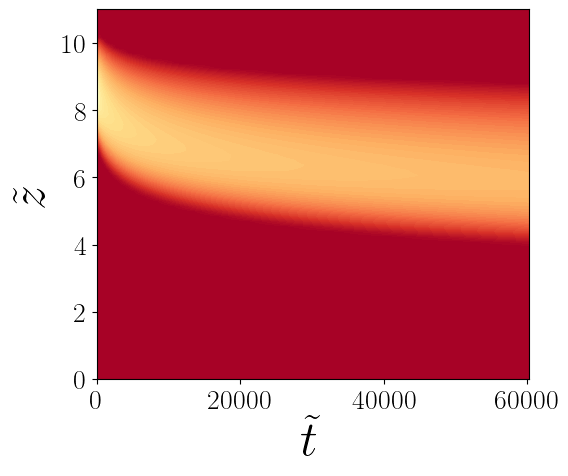}
		\caption{}\label{multi_phase_model}
	\end{subfigure}
	~ 
	\begin{subfigure}[t]{0.51\textwidth}
		\centering \includegraphics[width=\textwidth]{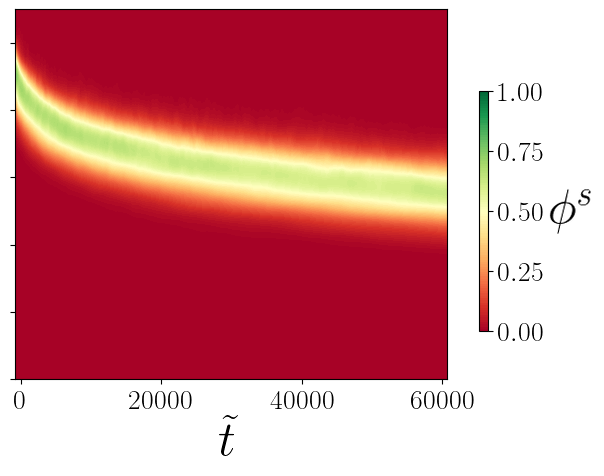}
		\caption{}\label{DEM}
	\end{subfigure}
	\caption{Spatio-temporal plot of the small particle concentration in the bed obtained with (a) the multi phase model, (b) the DEM simulation.}\label{spatio_temp_1}
\end{figure}
The system of partial differential equations \eqref{eq:MassConservation}-\eqref{MomSmallZ} is solved numerically for the configuration shown in figure \ref{numerical_set_up} with the initial concentration profile of figure \ref{concentration_init}. In these equations, the empirical segregation function $\mathcal{F(\mu)}$ is the one proposed in equation \eqref{revisited_Fmu} and the drag coefficient $c$ is equal to 3 as suggested by \citet{TripathiDensitydifferencedrivensegregation2013}.
Because the fluid is incompressible, there is no equation of state for the fluid pressure.
Nevertheless, remembering that $\epsilon+\Phi=1$ and defining the volume averaged velocities, $w  = \epsilon w^f+ \Phi  w^m$, it can be demonstrated that the particle-fluid mixture is incompressible. A PISO (Pressure Implicit with Splitting of Operators) algorithm classically developed for incompressible Navier-Stokes equations is used to solve the pressure-velocity coupling. 
As the fluid pressure $p^f$ is the sum of the hydrostatic pressure and of the excess pore presure $p^f = \overline{p^f} + \rho^f g z$, the PISO algorithm consists in solving the momentum balance equations without $\overline{p^f}$ in a predictor step. Then, using the predicted velocity fields, a Poisson equation is solved to find $\overline{p^f}$. Once the pressure is found, the velocity fields are corrected. This kind of algorithm has already been used to model sediment transport in \citet{ChauchatSedFoam23Dtwophase2017} and \citet{Chauchatcomprehensivetwophaseflow2018}.\\

\begin{figure}
	\centering
	\begin{subfigure}[b]{0.48\textwidth} 
		\centering \includegraphics[width=\textwidth]{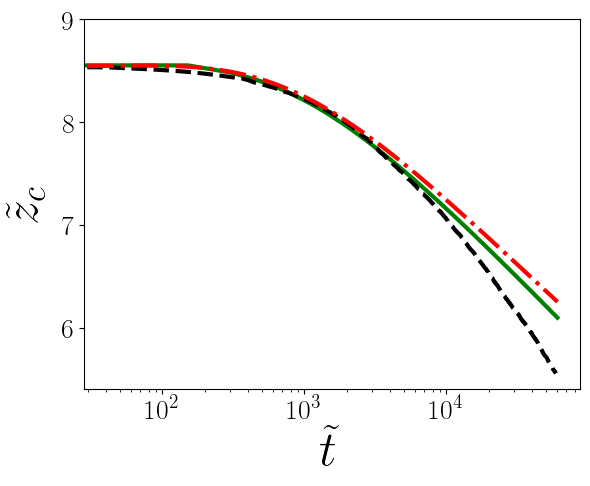}
		\caption{}\label{mass_center_1}
	\end{subfigure}
	\begin{subfigure}[b]{0.5\textwidth}
		\centering \includegraphics[width=\textwidth]{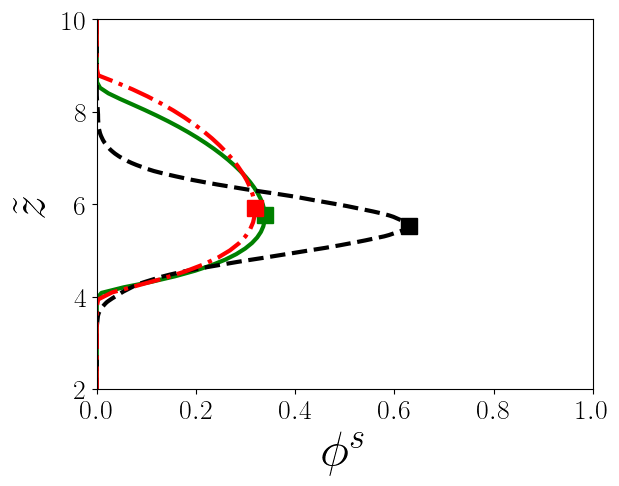}
		\caption{}\label{final_profile_1}
	\end{subfigure}
	\caption{Results of the simulation for the multi-phase flow model (\textcolor{C0}{\protect\tikz[baseline]{\protect\draw[line width=0.35mm, dash dot] (0,.5ex)--++(0.65,0) ;}}), the advection-diffusion model of \eqref{adv-diff} (\textcolor{C1}{\protect\tikz[baseline]{\protect\draw[line width=0.35mm] (0,.5ex)--++(0.65,0) ;}}) and the DEM ({\protect\tikz[baseline]{\protect\draw[line width=0.3mm,densely dashed] (0,.5ex)--++(0.65,0) ;}}). (a) Temporal evolution of the center of mass. (b) Concentration profiles of small particles at the end of the simulation ($\tilde{t} \simeq 60000$). {\protect\tikz[baseline]{\protect\draw[fill] (0.05,0.05) rectangle (0.2,0.2) ;}} represents the maximum concentration of the profiles.}\label{profiles_1}
\end{figure}

Figure \ref{spatio_temp_1} shows the results of the spatio-temporal evolution of the small particle concentration for the multi-phase flow model (figure \ref{multi_phase_model}) and for the DEM (figure \ref{DEM}). First, it can be seen that the dynamics predicted by the multi-phase flow model is similar to the DEM. The position of the bottom of the layer is about the same in both cases. More quantitatively, the center of mass $\tilde{z}_c$ of the small particle layer as a function of time is compared with the DEM in figure \ref{mass_center_1}.
After a first transient phase ($\tilde{t}>1\times 10^3$), the center of mass position is linear in the semilog plot, indicating that the logarithmic descent observed in the DEM simulation is well reproduced by the multi-phase flow model. The slope of the curve, representing coefficient $a_1$ of equation \eqref{log_descent_formula} is $0.68$ in the DEM simulation and $0.49$ for the multi-phase flow model, corresponding to an error of $28\%$. 
In addition, figure \ref{final_profile_1} shows that in both models, the bottom of the layer is positioned at the same depth indicating that the multi-phase model reproduces well the bottom controlled behaviour observed by \citet{Chassagne2020} with DEM simulations.
However the Gaussian-like profile is not reproduced by the multi-phase flow model and a wider profile is obtained. In figure \ref{multi_phase_model} the maximum concentration $max(\phi^s)$ (indicated by {\protect\tikz[baseline]{\protect\draw[fill] (0.05,0.05) rectangle (0.2,0.2) ;}} in figure \ref{final_profile_1}) is almost two times smaller than the one predicted by the DEM simulation, while the extent of the small particle layer is much larger. 
These results indicate that, with the current parametrisation the multi-phase flow model is relevant to qualitatively predict segregation dynamics. However, the error on the segregation velocity and the discrepancies on the concentration profile clearly show that the model needs to be improved.\\

\subsection{Evaluation of the advection-diffusion model}
\label{compare_models}
In order to determine the ability of the advection-diffusion model to reproduce the same results as the multi-phase flow model, equation \eqref{adv-diff} is solved numerically.
The resolution strategy is based on a conservative Godunov schemes where a no flux condition is applied at the bottom and on top of the vertical domain. A vertical discretisation of $d\tilde{z} = h / 80$ is taken and the time step is computed in order to satisfy the CFL condition. The initial solution is the same than in the multi phase flow (figure \ref{concentration_init}).

The numerical solution at time $\tilde{t} = 60000$ is plotted in figure \ref{profiles_1}. Both the center of mass (figure \ref{mass_center_1}) and the concentration profile (figure \ref{final_profile_1}) are almost superimposed with the multi-phase flow model solution, only slightly differing by numerical diffusion. Therefore, both models can be considered as strictly equivalent, meaning that the physical behaviour of the segregation forces added in the momentum equation of the small particles is accurately predicted by the single advection and diffusion coefficients of equation \eqref{adv-diff}. Moreover, when deriving the advection-diffusion equation, it was assumed that the mixture volume fraction $\Phi = cste$, that the vertical acceleration of the small particles, the vertical mixture velocity and that the fluid drag were negligible. The strong agreement between the models demonstrate that these assumptions are valid.

This new equation represents an important step in the upscaling since the behaviour of small particles can be predicted without solving the entire multi-phase flow model, providing a speed-up of one thousand for the numerical resolution of the equations. In the light of this result, the bidisperse segregation problem can be simply viewed as a competition between the advection coefficient $S_r$ and the diffusion coefficient $D$ of small particles. \\

\section{Discussion}
\label{discussion_part}
The multi-phase flow model and the associated advection-diffusion model are able to reproduce qualitatively the DEM simulations and the main properties of segregation in bedload transport obtained by \citet{Chassagne2020} (bottom controlled segregation, logarithmic descent of the small particles). However, the segregation velocity is lower than in the DEM simulation and the shape of the small particle concentration profile is not adequatly reproduced. So far, the inter-particle drag force and the segregation force from \citet{GuillardScalinglawssegregation2016} and  \citet{TripathiDensitydifferencedrivensegregation2013} have not been modified. Yet, these forces were proposed in very different configurations. Therefore, in section \ref{F_cste_part}, a sensitivity analysis to the empirical segregation function is presented. Then, in section \ref{Optimization}, new formulations of the empirical segregation function $\mathcal{F}$ and of the drag coefficient $c$ will be proposed based on the DEM results.\\
As the advection-diffusion equation results are strictly identical with the multi-phase flow equations, the discussion and the associated simulations will only be performed with the advection-diffusion equation.

\subsection{Investigation of the empirical segregation function $\mathcal{F}$} 
\label{F_cste_part}
Numerical data from \citet{GuillardScalinglawssegregation2016} for the empirical segrgeation function $\mathcal{F}(\mu)$ exhibit a significant scatter and it is possible to show that a constant function could also fit the data (see appendix \ref{AppendixA}). In this section, an analysis of the sensitivity to the empirical segregation function, taken as constant and varying from  $\mathcal{F} = 1$ to $\mathcal{F} = 15$, is presented.
\begin{figure}
	\begin{subfigure}[b]{0.51\textwidth} 
		\includegraphics[width=\textwidth]{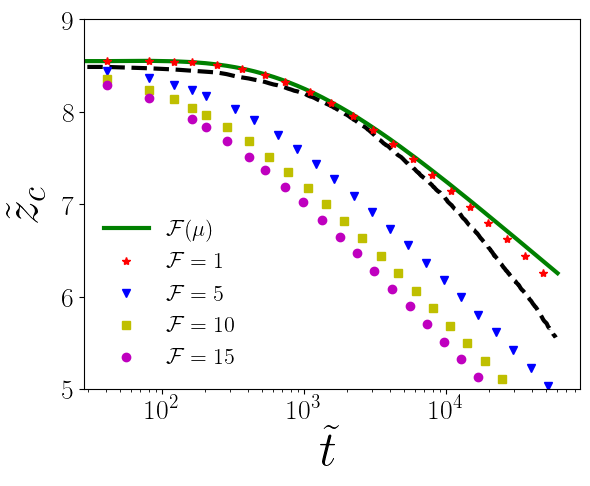}
		\caption{}\label{F_cste_a}
	\end{subfigure}
	\begin{subfigure}[b]{0.54\textwidth}
		\includegraphics[width=\textwidth]{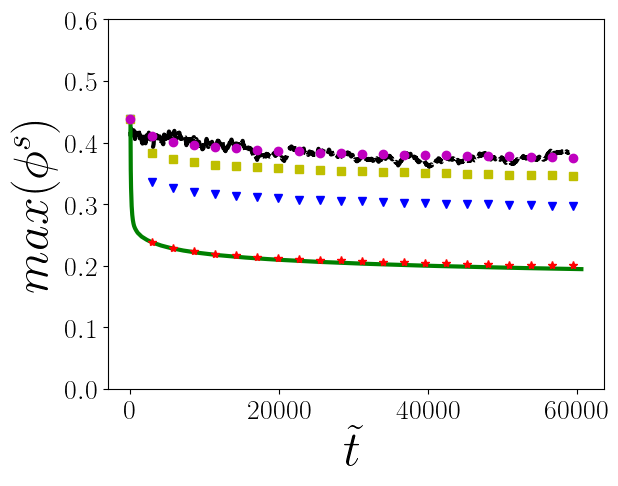}
		\caption{}\label{F_cste_b}
	\end{subfigure}
	\caption{Comparison with the DEM ({\protect\tikz[baseline]{\protect\draw[line width=0.3mm,densely dashed] (0,.5ex)--++(0.65,0) ;}}) for various values of $\mathcal{F} = cste$ for the temporal evolution of (a) the center of mass and (b) the maximum concentration of $\phi^s$.}\label{F_cste}
\end{figure}

Figure \ref{F_cste_a} shows the temporal evolution of the small particle center of mass for the different values of the empirical segregation function $\mathcal{F}$. The linear evolution in the semi-logarithmic plot is conserved with the same slope (coefficient $a_1$ in equation \ref{log_descent_formula}), whatever the value of the empirical segregation function, meaning that the segregation velocity $d\tilde{z}_c / d\tilde{t}$ is not modified when changing the empirical segregation function. Increasing $\mathcal{F}$ only makes the curves shift vertically.
Figure \ref{F_cste_b} shows that the maximum concentration is better predicted when increasing the empirical segregation function. When reaching $\mathcal{F} = 15$ agreement with DEM results is perfect. 
The previous simulation, where the empirical segregation function $\mathcal{F}$ is a function of the friction coefficient $\mu$, is also plotted in these figures. It is interesting to note that simulation with $\mathcal{F} = 1$ is almost superimposed with the one obtained with $\mathcal{F(\mu)}$. 
These observations tend to show that the friction coefficient dependency has a small influence on the size segregation configuration investigated. Therefore, taking $\mathcal{F}$ as constant is a good approximation, at first order.\\

\citet{Chassagne2020} showed that the advection coefficient $S_r$ was a function of the inertial number $I$ (see equation \ref{SegregationNumber}) and linked the logarithmic descent to the exponential form of $I$ in $S_r$.
In the proposed model, the advection coefficient $S_r = \mathcal{F} St^p \partial \tilde{p}^m / \partial \tilde{z}$ (equation \ref{SegregationNumber_DiffusionCoefficient}). In this coefficient,  since the solid pressure gradient is constant in the bed (equation \ref{hydrostatic_pressure_gradient}) and $\mathcal{F} = cste$, there is only one non-constant variable which is the inverse of the granular viscosity $1/\eta^p$ appearing in the granular Stokes number (equation \ref{particle_stokes_number}). As figure \ref{F_cste_a} shows that the logarithmic descent is still preserved with this parametrisation, it demonstrates that the viscosity profile is mainly responsible for the logarithmic descent. Therefore, in the proposed advection coefficient $S_r$, the empirical segregation function $\mathcal{F}$ controls the strength of the segregation force and can be seen as a forcing parameter. In contrast, the velocity at which the small grains are descending is controlled by the granular viscosity in the granular Stokes number $St^p$. In this way, the segregation problem can simply be seen as the settling of small particles under a force proportional to the empirical segregation function $\mathcal{F}$, into a complex fluid having a variable viscosity.

It should be noted that there is a direct relation between the dimensionless granular viscosity and the inertial number (see appendix \ref{AppendixC}) written
\begin{equation}
I = \dfrac{\mu \sqrt{\tilde{p}^m}}{\tilde{\eta^p}},
\label{eta_p_I}
\end{equation}
where $\tilde{\eta}^p = \eta^p / \rho^p d_l W$ is the dimensionless granular viscosity and $\mu = \tilde{\tau}_{xz}^m / \tilde{p}^m$ is the friction coefficient.
Since the variation of $\mu$ and $\sqrt{\tilde{p}^m}$ is small compared with the exponential profile of $\tilde{\eta}^p$, the inversely proportional relation between the granular viscosity and the inertial number shows that the inertial number dependency found by \citet{Chassagne2020} can be seen as a dependency on the inverse granular viscosity, confirming the important role of the granular viscosity in the segregation dynamics.\\

Figure \ref{F_cste_b} showed that a better maximum concentration was predicted with $\mathcal{F} = 15 $. However, the value of the empirical segregation function $\mathcal{F}$ should have no influence on the diffusion as $D = \phi^s / \Phi \tilde{p}^m St^p$. This influence can be explained by the P\'eclet number $Pe =S_r/D$. Indeed, when $\mathcal{F}$ increases, the advection coefficient $S_r$ increases as well and makes the P\'eclet number higher. As a result the diffusive effect becomes small compared with the advection and thus, the small particles stay more concentrated.
However, the dynamics of the center of mass shows a better agreement for $\mathcal{F}=1$ (see figure \ref{F_cste_a}) than for $\mathcal{F} = 15 $, meaning that the diffusion coefficient needs to be improved.

\begin{figure}
	\begin{subfigure}[b]{0.52\textwidth} 
		\includegraphics[width=\textwidth]{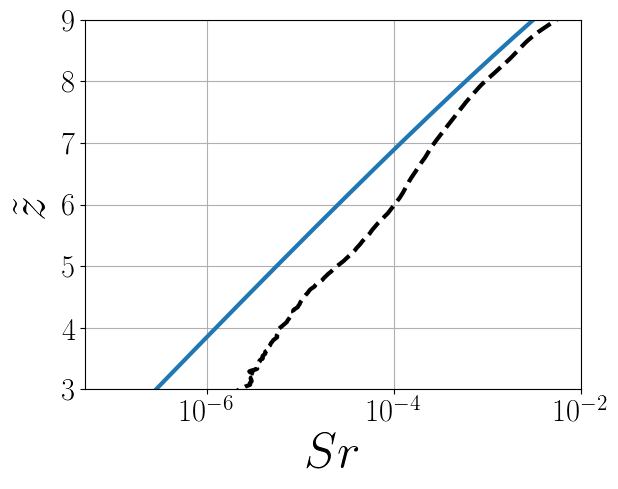}
		\caption{}\label{compare_Sr}
	\end{subfigure}
	\begin{subfigure}[b]{0.478\textwidth} 
		\includegraphics[width=\textwidth]{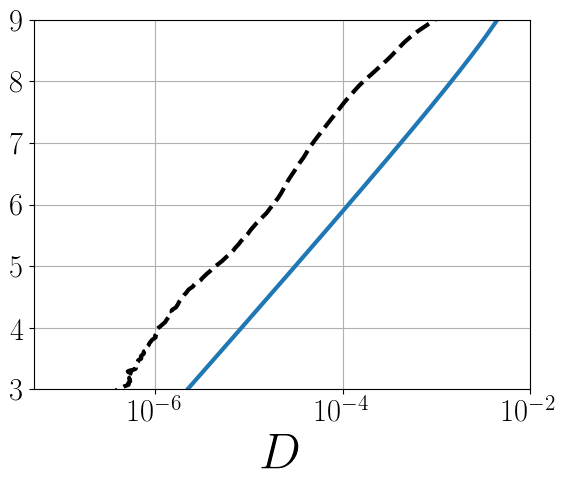}
		\caption{}\label{compare_D}
	\end{subfigure}
	\caption{(a) Advection coefficient in the bed (equation \ref{adv-diff}) with $\mathcal{F} = 1$, for $N_s = 1.5$ (\textcolor{C2}{\protect\tikz[baseline]{\protect\draw[line width=0.35mm] (0,.5ex)--++(0.65,0) ;}}) and  $S_{r0} I^{0.85}$ proposed by \citet{Chassagne2020} ({\protect\tikz[baseline]{\protect\draw[line width=0.35mm,densely dashed] (0,.5ex)--++(0.65,0) ;}}). (b) Diffusion coefficient in the bed from equation \eqref{adv-diff} for $\phi^s = max(\phi^s)$ for the case $N_s=1.5$  (\textcolor{C2}{\protect\tikz[baseline]{\protect\draw[line width=0.35mm] (0,.5ex)--++(0.65,0) ;}}) and $D_0 I^{0.85}$ proposed by \citet{Chassagne2020} ({\protect\tikz[baseline]{\protect\draw[line width=0.35mm,densely dashed] (0,.5ex)--++(0.65,0) ;}}).}\label{figure7}	
\end{figure}

The advection coefficient $S_r$ and the diffusion coefficient $D$ have been plotted in figures \ref{compare_Sr} and \ref{compare_D} with $\mathcal{F} = 1$, for $\phi^s = max(\phi^s)$.
For $\tilde{z} > 7$, $S_r$ is close to the values predicted by the DEM. Under this limit, both curves are exponentially decreasing into the bed but with a different slope leading to discrepancies. This shows that the $1/\eta^p$ dependency has a fundamental role in the vertical structure of the advection coefficient $S_r$. Yet, another dependency with depth is probably missing in the empirical segregation function or in the drag coefficient $c$ to find again a similar slope to the DEM.
Surprisingly, the proposed diffusion coefficient has the same slope as the one predicted by the DEM,  which means that it contains the correct depth dependency. However, its value is too high by a factor ten, explaining why the advection-diffusion results are too diffusive. 


\subsection{Missing dependencies in the particle-scale forces}
\label{Optimization}
In the present paper, the advection and diffusion coefficients (equation \ref{SegregationNumber_DiffusionCoefficient}) have been derived from particle-scale segregation and solid drag forces of \citet{GuillardScalinglawssegregation2016} and \citet{TripathiDensitydifferencedrivensegregation2013}. However, figure \ref{figure7} shows that both coefficients should be improved so as to match the DEM results. The segregation and solid drag forces of \citet{GuillardScalinglawssegregation2016} and \citet{TripathiDensitydifferencedrivensegregation2013} have been established in idealised granular segregation configurations (e.g. unique intruder, simple forcing, 2D DEM) so that one can expect the two formulations to miss some dependencies when considering more general cases (mixture of small and large particles, 3D modelling or complex forcing). 
Such dependencies probably lie in the drag coefficient $c$ and the empirical segregation function $\mathcal{F}$ contained in the advection and diffusion coefficients. While the drag coefficient $c$ was taken constant in their study, \citet{TripathiDensitydifferencedrivensegregation2013} suggested that it could depend on the local concentration of particles. Similarly to an hindrance function in a fluid flow, one indeed expects an increase of the effective solid drag force on a particle with increasing concentration. This dependency of the drag coefficient in the local particle concentration should also impact the diffusion coefficient profile (see figure \ref{compare_D}) and makes it possible to match the DEM.
In addition, to correct the slope of the advection coefficient $S_r$ (see figure \ref{compare_Sr}), only the empirical segregation function $\mathcal{F}$ should vary with depth. 

As detailed in section \ref{Summary_DEM_part}, \citet{Chassagne2020} have been able to express the advection and diffusion coefficients dependencies on the inertial number $I$ (see equations \ref{SegregationNumber} and \ref{DiffusionCoefficient_DEM}). In the following, both DEM and advection-diffusion coefficients are compared so as to extract the potential missing dependencies of $\mathcal{F}$ and $c$ from the DEM coefficients and to propose new formulations of these parameters. Then, it is verified that the results from the advection-diffusion model are consistent when using these proposed closures.\\

In order to compare the advection and diffusion coefficients to the DEM and to find the missing dependencies, it is first shown that the coefficients of equation \eqref{SegregationNumber_DiffusionCoefficient} can be expressed with an inertial number dependency as in the DEM. Indeed, as already shown in equation \eqref{eta_p_I}, the granular Stokes number can be rewritten as a function of the inertial number $I$:
\begin{equation}
St^p = \dfrac{I}{6c \mu \sqrt{\tilde{p}^m}}.
\label{Stokes_I}
\end{equation}
With this new definition, the advection and diffusion coefficients obtained in equation \eqref{SegregationNumber_DiffusionCoefficient} can be rewritten as a function of the inertial number $I$:
\begin{equation}
S_r = \dfrac{I \mathcal{F}}{6 c \mu \sqrt{\tilde{p}^m}} \dd{\tilde{p}^m}{\tilde{z}}  \quad \text{and} \quad D = \dfrac{ \phi^s \sqrt{\tilde{p}^m} I}{\Phi 6 c \mu}.
\label{SegregationNumber_DiffusionCoefficient_2}
\end{equation}
One can notice that $\sqrt{\tilde{p}^m} I = \tilde{\dot{\gamma}}$, which makes the diffusion coefficient directly proportional to the shear rate. Such a dependency for the diffusion coefficient was found in a size bidisperse case by \citet{CaiDiffusionsizebidisperse2019} with DEM simulations. This shows that the particle-particle force based model is able to explain a dependency found in a different experiment. It represents a powerful argument to show the robustness of the proposed model. Yet, the power $0.85$ highlighted by the DEM does not seem to appear.
\begin{figure}
	\centering\includegraphics[width=0.6\textwidth]{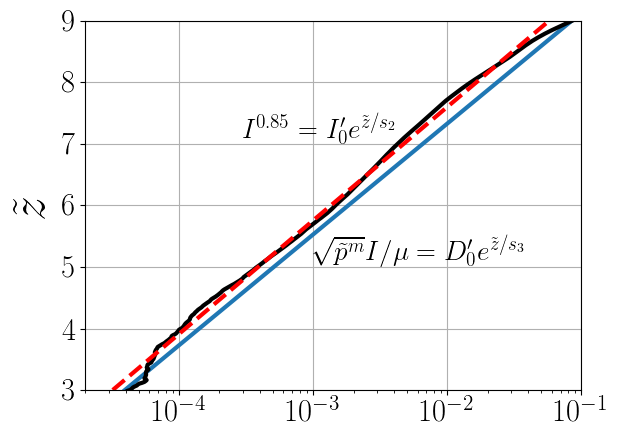}
	\caption{$I^{0.85}$ computed from DEM results ({\protect\tikz[baseline]{\protect\draw[line width=0.35mm] (0,.5ex)--++(0.65,0) ;}}) and its exponential fit (\textcolor{C0}{\protect\tikz[baseline]{\protect\draw[line width=0.35mm,densely dashed] (0,.5ex)--++(0.65,0) ;}}) compared with $\sqrt{\tilde{p}^m} I / \mu$ (\textcolor{C2}{\protect\tikz[baseline]{\protect\draw[line width=0.35mm] (0,.5ex)--++(0.65,0) ;}}).  In both fits, $I_0^{\prime} = 7.52 \times 10^{-7}$, $s_2 = 0.8$, $D_0^{\prime} = 8.46 \times 10^{-7}$ and $s_3 = 0.78$.}\label{compare_I_I085}
\end{figure}
In figure \ref{compare_I_I085} the profile of $\sqrt{\tilde{p}^m} I / \mu$, the profile of $I^{0.85}$ from the DEM and from exponential fitting are plotted. One can observe that values only differ by a factor $D_0^{\prime} / I_0^{\prime} = 1.13$ and that the exponential evolution with depth is the same, which proves that in this case
\begin{equation}
\dfrac{\sqrt{\tilde{p^m}} I}{\mu} \propto I^{0.85}.
\label{diff_propo_to_I085}
\end{equation} 
Such a result shows that the diffusion coefficient of the proposed model (equation \ref{SegregationNumber_DiffusionCoefficient_2}) has the same dependency with $I^{0.85}$ as the diffusion coefficient proposed by \citet{Chassagne2020}. This explains the identical evolution with depth between both coefficients in figure \ref{compare_D}, and shows that the advection equation and the multi-phase flow model are physically consistent.

Assuming that the drag coefficient includes the accurate dependencies, the following equality should be obtained between diffusion coefficients of the DEM and the one proposed in equation \eqref{SegregationNumber_DiffusionCoefficient_2}:
\begin{equation}
\dfrac{ \phi^s \sqrt{\tilde{p^m}} I}{\Phi 6 c \mu} = D_0I^{0.85}
\label{diff}
\end{equation} 
As demonstrated in equation \eqref{diff_propo_to_I085}, $\sqrt{\tilde{p}^m} I / \mu I^{0.85} = C_0$, where $C_0$ is a constant (see appendix \ref{AppendixD} for the details) the drag coefficient $c$ can be deduced from equation \eqref{diff}:
\begin{equation}
c(\phi^s) = \dfrac{C_0}{6\Phi D_0} \phi^s = 31 \phi^s.
\label{drag_1}
\end{equation}
As a consequence of this linear scaling in concentration, the dependency on the small particle volume fraction vanishes in the diffusion coefficient (equation \ref{SegregationNumber_DiffusionCoefficient}). This modification will tend to smooth out the small particle concentration profile. In this drag coefficient, when $\phi^s \rightarrow 0$ (i.e. one small particle in a bath of large particles), the drag coefficient vanishes while it should reach a constant value, $c=3$ as shown by \citet{TripathiDensitydifferencedrivensegregation2013}.
To ensure a consistent formulation, it is therefore proposed $c(\phi^s)$ to read
\begin{equation}
c(\phi^s) = 28 \phi^s + 3,
\label{cphis_2}
\end{equation}
which tends to $3$ when $\phi^s \rightarrow 0$ and to $31$ when $\phi^s \rightarrow 1$.\\

As mentioned in the last section, the empirical segregation function $\mathcal{F}$ is expected to depend on depth (see figure \ref{compare_Sr}). In this way, the advection coefficient should correspond to the DEM and it follows
\begin{equation}
\dfrac{I \mathcal{F}}{6 c(\phi^s) \mu \sqrt{\tilde{p}^m}} \vert\dd{\tilde{p}^m}{\tilde{z}}\vert =S_{r0}I^{0.85}.
\label{adv}
\end{equation}
From this equation, the empirical segregation function is easily expressed as
\begin{equation}
\mathcal{F}(\mu, \tilde{p}^m, \phi^s) = \dfrac{6 S_{r0} c(\phi^s) \rho^p}{\Phi (\rho^p - \rho^f)}  \mu \sqrt{\tilde{p}^m} I^{-0.15},
\label{F_1}
\end{equation}
Equation \eqref{F_1} makes it possible to express the missing dependencies in the empirical segregation function $\mathcal{F}$. One can note that a dependency with the friction coefficient is found as predicted by \citet{GuillardScalinglawssegregation2016}. In addition, the empirical segregation function is found to depend on the small particle concentration. The dependency on $\sqrt{\tilde{p}^m} I^{-0.15}$ exhibits an additional more complex mechanism in the segregation force. \\

With the new formulations of the solid drag coefficient $c(\phi^s)$ and the empirical segregation function $\mathcal{F}(\mu, \tilde{p}^m, \phi^s)$, it is verified that the DEM results can be reproduced. This is done with simulations accounting for different initial quantities of small particles ($N_s = {0.5, 1, 1.5, 2}$).
\begin{figure}
	\centering
	\centering \includegraphics[width=\textwidth]{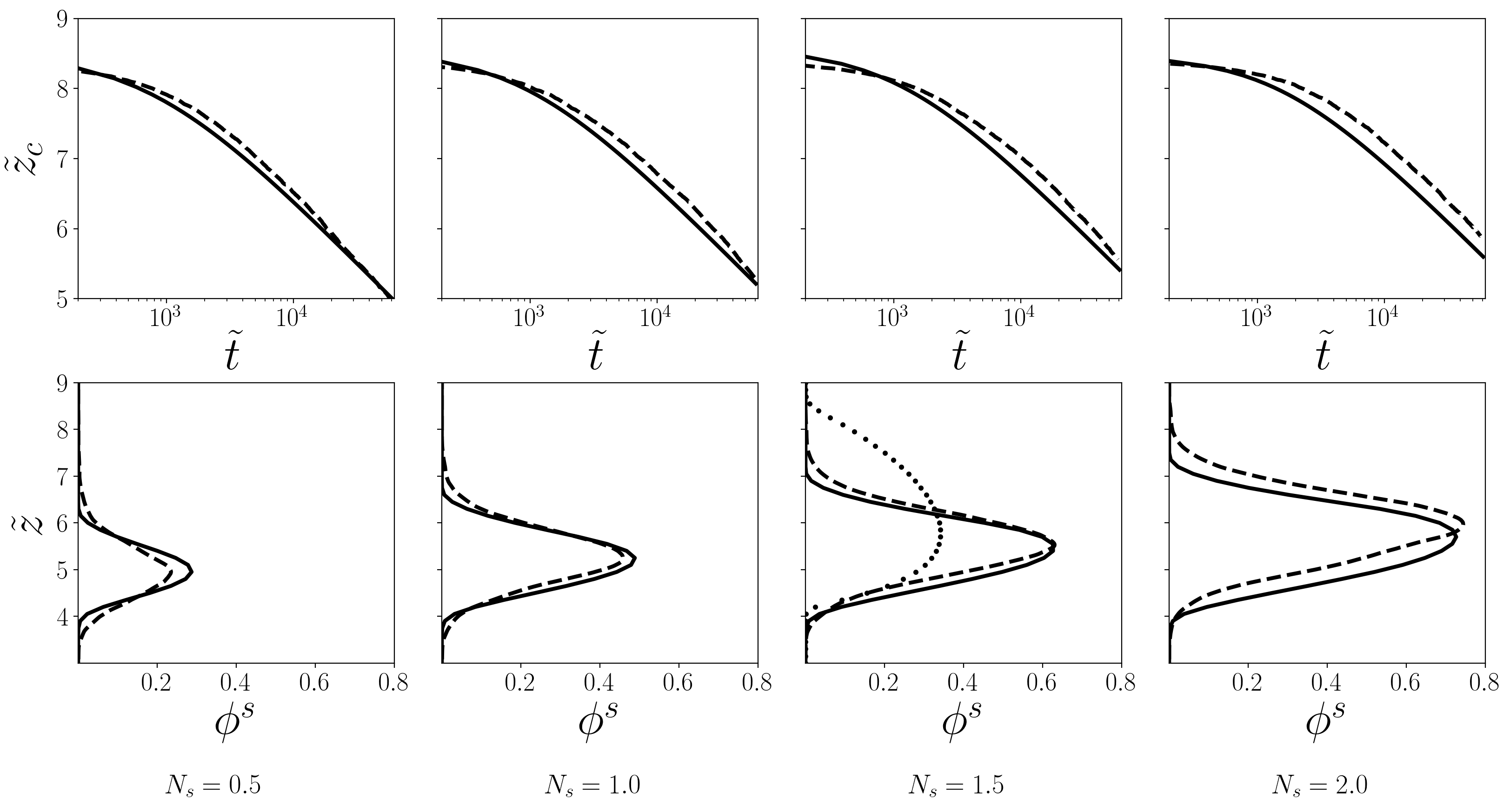}
	\caption{Upper part of the panel: temporal evolution of the center of mass for $N_s = {0.5, 1, 1.5, 2}$. Lower part of the panel: final small particle concentration profile for $N_s = {0.5, 1, 1.5, 2}$. {\protect\tikz[baseline]{\protect\draw[line width=0.35mm,densely dashed] (0,.5ex)--++(0.65,0) ;}} are the DEM results from \citet{Chassagne2020}. The concentration profile obtained without any parametrisation (from figure \ref{final_profile_1}) has also been plotted ({\protect\tikz[baseline]{\protect\draw[line width=0.35mm,dotted] (0,.5ex)--++(0.65,0) ;}}).\label{results_c_phis}}
\end{figure}
Figure \ref{results_c_phis} shows the results for the time evolution of the center of mass and the final concentration profile. Both the mass centers and the final concentration profiles are fairly well superimposed with the DEM results. Therefore, the new parametrisation is consistent with the DEM simulations.
The concentration profile with the original parametrisations of $c$ and $\mathcal{F}(\mu)$ (see figure \ref{final_profile_1}) has also been plotted in figure \ref{results_c_phis} ($N_s=1.5$). The shape of the final concentration profile has drastically changed, from a bell-shape to the expected Gaussian-like shape. This is attributed to the small particle concentration dependency in the drag coefficient, which cancels the original concentration dependency in the diffusion coefficient.

Lastly, note that changing the parametrisation of the forces still yields physical solutions which proves that the advection-diffusion model and the corresponding multi-phase flow model are physically consistent and robust.\\


\subsection{Influence of the size ratio}
\citet{GuillardScalinglawssegregation2016} showed that the segregation force depends on the size ratio $r = d_l / d_s$ and exhibits a maximum for a size ratio of $r = 2$. Based on DEM simulations, \citet{Chassagne2020} also studied the size-ratio dependency and found the segregation velocity of the percolating small particles to be a monotonic increasing function of $r$. Best fit of the DEM results suggested the following dependency $f(r) = 0.45\left(e^{(r-1)/1.59}-1\right)$ for the advection coefficient $S_r$.\\

\begin{figure}
	\centering
	\centering \includegraphics[width=\textwidth]{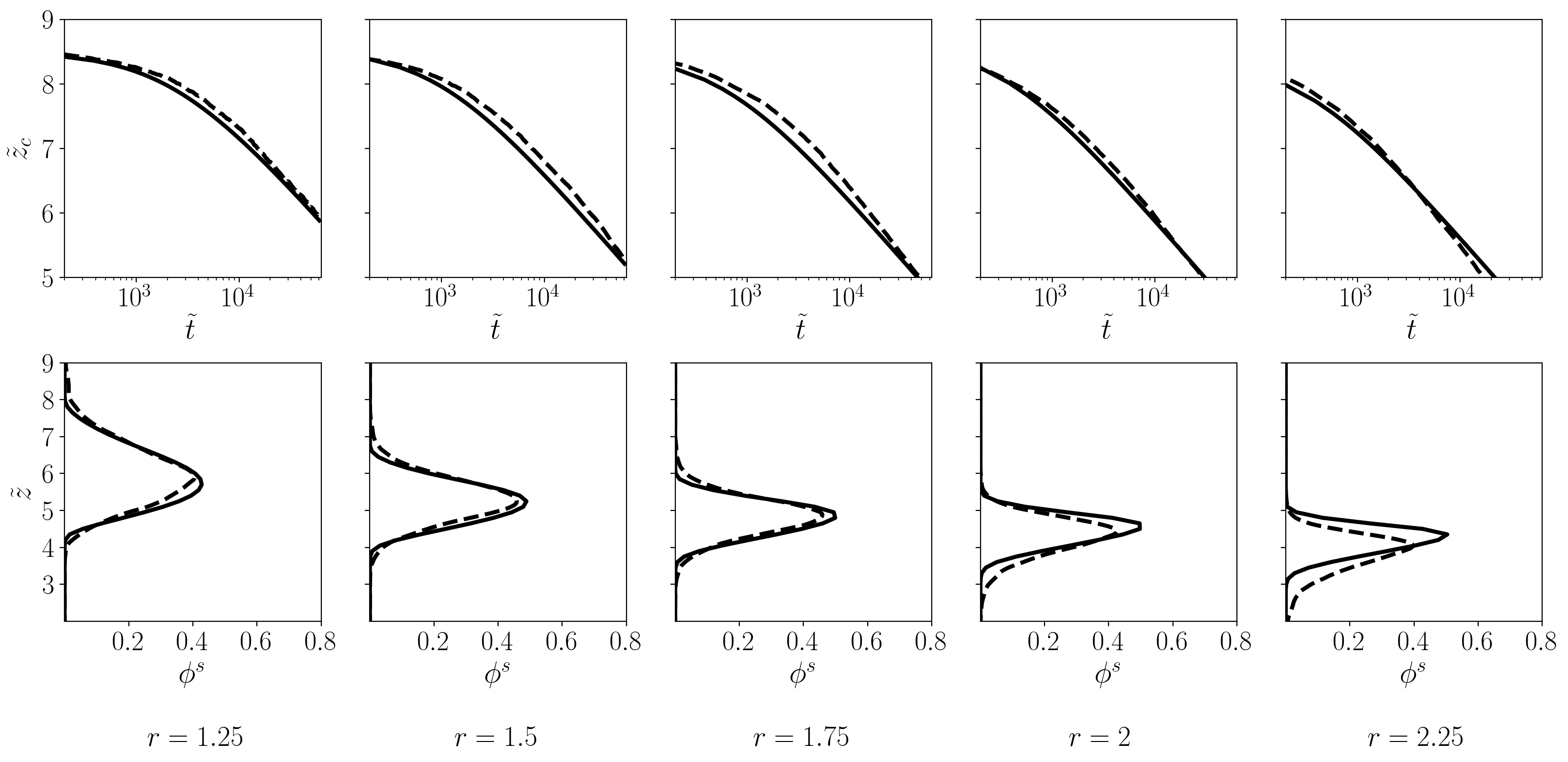}
	\caption{Upper part of the panel: temporal evolution of the center of mass for $N_s = 1$ and $r = {1.25, 1.5, 1.75, 2, 2.25}$. Lower part of the panel: final small particle concentration profile for $N_s = 1$ and $r = {1.25, 1.5, 1.75, 2, 2.25}$. In these figures, {\protect\tikz[baseline]{\protect\draw[line width=0.35mm] (0,.5ex)--++(0.65,0) ;}} corresponds to the advection-diffusion model and {\protect\tikz[baseline]{\protect\draw[line width=0.35mm,densely dashed] (0,.5ex)--++(0.65,0) ;}} are the DEM results from \citet{Chassagne2020}.\label{size_ratio_2}}
\end{figure}

This dependency is introduced into the empirical segregation function $\mathcal{F}$ as follows
\begin{equation}
\mathcal{F}(\mu, \tilde{p}^m, \phi^s, r) = f(r) \dfrac{6 S_{r0} c(\phi^s) \rho^p}{\Phi (\rho^p - \rho^f)} \mu \sqrt{\tilde{p}^m} I^{-0.15}.
\end{equation}
Using this parametrisation, simulations have been performed for $r={1.25, 1.5, 1.75, 2, 2.25}$. The results are plotted in figure \ref{size_ratio_2} and compared with DEM simulations.
For each case, the center of mass position is in very good agreement with the DEM simulations. For the lower size ratio, the concentration profiles are superimposed with the DEM results while for the higher size ratios ($r=2$ and $r = 2.25$) the maximum concentrations are slightly higher than the DEM. This indicates that the model is not diffusive enough.
Indeed, the size ratio dependency has only been introduced in the advection coefficient. As shown in section \ref{F_cste_part}, the shape of the concentration profile results from a subtle balance between advection and diffusion through the P\'eclet number $Pe$. For the highest size ratio, this balance is not perfectly reproduced by the proposed model, which indicates that the diffusion coefficient should also depend on the size ratio. This would imply that the granular Stokes number also depends on the size ratio. It could explain why \citet{GuillardScalinglawssegregation2016} found a maximum segregation force for a size ratio $r=2$, while \citet{Chassagne2020} found the advection coefficient $S_r$ to increase exponentially with the size ratio. 
Further research is needed to elucidate this point through a detailed investigation of the granular Stokes number dependency on the size ratio.

\section{Conclusion}

In this contribution size segregation in bidisperse systems has been investigated with special focus on bedload transport. The originality of the work presented herein is to propose a new multiphase flow approach, derived from a volume averaging technique, based on the most recent advances on particle-particle forces, namely the segregation force or buoyancy force from \citet{GuillardScalinglawssegregation2016} and the drag force from \citet{TripathiDensitydifferencedrivensegregation2013}. The proposed multiphase flow model formulation is very general and it can be applied to any immersed granular flow configuration. In a subsequent step, following the same procedure as in \cite{Thorntonthreephasemixturetheory2006}, an advection-diffusion model is derived from the multi-phase flow equations. This derivation makes it possible to identify the dependencies of the advection and diffusion coefficients with the local physical parameters of the flow such as the volume fraction of small particles, the mixture granular pressure and its gradient, the granular Stokes number and the segregation parameter. 

Both models have been tested against the Discrete Element Model simulations of \citet{Chassagne2020} for bidisperse turbulent bed load transport. Without any tuning of the forces from \citet{GuillardScalinglawssegregation2016} and \citet{TripathiDensitydifferencedrivensegregation2013}, both continuum models qualitatively reproduce the main features of size segregation. This demonstrates that the scaling of the advection coefficient with the inertial number observed by \citet{FryEffectpressuresegregation2018} and \citet{Chassagne2020} can be explained thanks to the dependency of the advection coefficient on the granular Stokes number and the underlying presence of the granular viscosity.
Using the discrete element simulation results, improved parametrisations for the advection and diffusion coefficients have been proposed. They suggest that  the empirical segregation function from \citet{GuillardScalinglawssegregation2016} and the drag coefficient from \citet{TripathiDensitydifferencedrivensegregation2013} should incorporate a dependency on the small particle concentration. At last, the influence of the size ratio has been investigated and a dependency of the segregation function on the size ratio has been proposed.

In terms of perspectives for granular flows, the continuum models proposed herein are very general and should as well apply to dense dry granular flows. These models represent a general framework for developing and testing improved parametrisations for the segregation and granular drag forces in different flow configurations. Further work is needed to identify and propose more robust concentration, depth and size ratio dependencies of the empirical dimensionless coefficients appearing in both granular forces.  
Concerning the upscaling of size segregation processes in sediment transport applications two routes are opened. The first one consists in implementing the proposed multi-phase flow model in a 3D numerical model, such as sedFOAM \citep{ChauchatSedFoam23Dtwophase2017}, for the simulation of size segregation in complex sediment transport application such as riverbed armouring \citep{FreyHowRiverBeds2009}, scour around an hydraulic structure \citep{NAGEL2020} or wave-driven sediment transport involving sand mixtures \citep{ODONOGHUE2004}. The second route would be to couple the advection-diffusion model with a shallow water model for the fluid flow. Such a model would make it possible to address size segregation at the reach scale while taking into account granular scale processes in a physically consistent way.\\

Declaration of Interests. The authors report no conflict of interest

\section*{Acknowledgements}
This research was funded by the French Agence nationale de la recherche, project ANR-16-CE01-0005 SegSed 'size segregation in sediment transport'. The authors acknowledge the support of Irstea (now INRAE, formerly Cemagref). INRAE, ETNA is member of Labex TEC21 (Investissements d’Avenir Grant Agreement ANR-11-LABX-0030) and Labex Osug@2020 (Investissements d’Avenir Grant Agreement ANR-10-LABX-0056).

\appendix

\section{New formulation of $\mathcal{F(\mu)}$ for a single large particle}
\label{AppendixA}
In this part, it is shown that the empirical segregation function $\mathcal{F(\mu)}$ does not satisfy the dimensionless equation of the large intruder. Therefore, a new formulation is proposed.

\begin{figure}
	\centering{\includegraphics[scale=0.54]{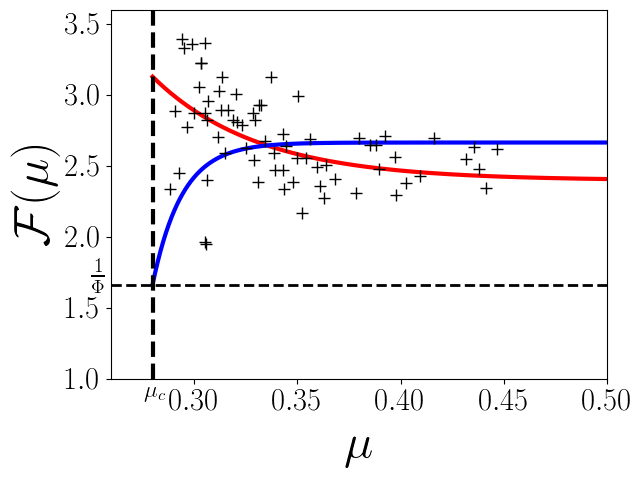}}
	\caption{Empirical segregation function $\mathcal{F(\mu)}$ of the segregation force $f_{seg} = V_l \mathcal{F}(\mu) \dd{P^s}{z}$ found by \citet{GuillardScalinglawssegregation2016} as a function of the local friction coefficient $\mu$. $+$ are the simulations found by \citet{GuillardScalinglawssegregation2016} using DEM. \textcolor{C0}{\protect\tikz[baseline]{\protect\draw[line width=0.35mm] (0,.5ex)--++(0.65,0) ;}} is the function they proposed and \textcolor{C2}{\protect\tikz[baseline]{\protect\draw[line width=0.35mm] (0,.5ex)--++(0.65,0) ;}} is $\mathcal F(\mu) = \dfrac{1}{\Phi} + \big(1-e^{-70(\mu-\mu_c)}\big)$, the improved function proposed.}
	\label{Fmu_plot}
\end{figure}

Considering an immobile bed, segregation should stop and therefore $w^l = 0$. In this case the drag forces vanish and the friction coefficient should be equal to the static friction coefficient $\mu = \mu_c$.
Therefore equation \eqref{dimensionless_lagrangian_eq} becomes
\begin{equation}
\dfrac{\rho^p-\rho^f}{\rho^p} + \mathcal{F}(\mu_c) \dd{\tilde{P^s}}{\tilde{z}} =0.
\label{lagrangian_eq_rest}
\end{equation}
For a dense flow with small velocity fluctuations, the small particle pressure is the lithostatic pressure $P^s = \Phi_{max} (\rho^p - \rho^f) g (h-z)$. The dimensionless particle pressure gradient is therefore $\partial \tilde{P^s} /\partial \tilde{z} = - \Phi_{max} (\rho^p - \rho^f)/\rho^p$. Introducing this expression in equation \eqref{lagrangian_eq_rest}, it finally imposes that
\begin{equation}
\mathcal{F}(\mu_c) = \dfrac{1}{\Phi_{max}}.
\label{Seg_Force_hydrostatic}
\end{equation}
The condition \ref{Seg_Force_hydrostatic} shows that the segregation force should balance the hydrostatic particle pressure, at rest. However, the functional form \eqref{function_Guillard} proposed by \citet{GuillardScalinglawssegregation2016} and plotted in figure \ref{Fmu_plot}, do not satisfy condition \eqref{Seg_Force_hydrostatic} in the quasi-static regime ($\mu<0.3$). In order to verify condition \eqref{Seg_Force_hydrostatic}, the following form is proposed:
\begin{equation}
\mathcal F(\mu) = \dfrac{1}{\Phi_{max}} + \big(1-e^{-70(\mu-\mu_c)}\big),
\label{revisited_Fmu_appendix}
\end{equation}
and is plotted in figure \ref{Fmu_plot}. It can be observed that the new formulation \eqref{revisited_Fmu} is close to equation \eqref{function_Guillard} for $\mu>>\mu_c$ but decreases to $1/\Phi_{max}$ in the static regime. This makes it possible to write the segregation force as
\begin{equation}
f_{seg} = \left(\dfrac{1}{\Phi_{max}} + \big(1-e^{-70(\mu-\mu_c)}\big)\right) \dd{\tilde{P^s}}{\tilde{z}}
\label{segregtion_force_appendix}
\end{equation}
The formulation \eqref{segregtion_force_appendix} shows that the term $1/\Phi_{max} \partial{\tilde{P^s}}/\partial{\tilde{z}}$ balances the hydrostatic pressure while the term
\begin{equation}
\big(1-e^{-70(\mu-\mu_c)}\big)\dd{\tilde{P^s}}{\tilde{z}}
\label{segregation_part}
\end{equation}
models an additional buoyancy effect as soon as the medium is sheared (i.e. $\mu>>\mu_c$). As a consequence, it appears to be the proper segregation force.

\section{Formulation of $\mathcal{F(\mu)}$ in the multi-phase flow case}
\label{AppendixB}
It was found that the empirical segregation function should be $\mathcal{F}(\mu) = \big(1-e^{-70(\mu-\mu_c)}\big)$, in the multi-phase flow model. An explanation is given thereafter.\\ 
Using  \eqref{continuousSmallLargeDrag} and \eqref{partialPressure}, equations \eqref{MomLargeZ} and \eqref{MomSmallZ} can be written as
\begin{multline}
\rho^p \left[\dd{\Phi^s w^s}{t}  +  \dd{\Phi^s w^s w^s}{z}\right] = -\dfrac{\Phi^s}{\Phi}\dd{p^m}{z} -\dfrac{p^m}{\Phi}\dd{\Phi^s}{z} - \Phi^s \dd{p^f}{z} - \Phi^s \rho^p g \cos \theta + \dfrac{\rho^p \Phi^s}{t_s}\left(w^f-w^s\right) \\- \dfrac{\rho^p \Phi}{t_{ls}} \big(w^s-w^m\big) + \Phi^l \mathcal{F}(\mu) \dd{p^m}{z}
\label{MomSmallZ_2_appendix}
\end{multline}	
and
\begin{multline}
\rho^p \left[\dd{\Phi^l w^l}{t}  +  \dd{\Phi^l w^l w^l}{z}\right] = -\dfrac{\Phi^l}{\Phi}\dd{p^m}{z} -\dfrac{p^m}{\Phi}\dd{\Phi^l}{z} - \Phi^l \dd{p^f}{z} - \Phi^l \rho^p g \cos \theta + \dfrac{\rho^p \Phi^l}{t_l}\left(w^f-w^l\right) \\+ \dfrac{\rho^p \Phi}{t_{ls}} \big(w^l-w^m\big) - \Phi^l \mathcal{F}(\mu) \dd{p^m}{z}.
\label{MomLargeZ_2_appendix}
\end{multline}	
Because $\partial{p^m}/\partial {z} = -\Phi \left(\rho^p - \rho^f\right) g \cos \theta$ equations \eqref{MomSmallZ_2_appendix} and \eqref{MomLargeZ_2_appendix} become
\begin{multline}
\rho^p \left[\dd{\Phi^s w^s}{t}  +  \dd{\Phi^s w^s w^s}{z}\right] = -\dfrac{p^m}{\Phi}\dd{\Phi^s}{z} + \dfrac{\rho^p \Phi^s}{t_s}\left(w^f-w^s\right) \\- \dfrac{\rho^p \Phi}{t_{ls}} \big(w^s-w^m\big) + \Phi^l \mathcal{F}(\mu) \dd{p^m}{z}
\label{MomSmallZ_3_appendix}
\end{multline}	
and
\begin{multline}
\rho^p \left[\dd{\Phi^l w^l}{t}  +  \dd{\Phi^l w^l w^l}{z}\right] = -\dfrac{p^m}{\Phi}\dd{\Phi^l}{z} + \dfrac{\rho^p \Phi^l}{t_l}\left(w^f-w^l\right) \\+ \dfrac{\rho^p \Phi}{t_{ls}} \big(w^l-w^m\big) - \Phi^l \mathcal{F}(\mu) \dd{p^m}{z},
\label{MomLargeZ_3_appendix}
\end{multline}	
where the hydrostatic pressure vanishes. As a consequence, there is no need for the segregation force mentioned in \eqref{segregtion_force_appendix} to contain the term $1/\Phi_{max}$ that balanced the hydrostatic pressure (see appendix \ref{AppendixA}).
For the multi-phase flow model, the segregation force is therefore
\begin{equation}
f_{seg} = \big(1-e^{-70(\mu-\mu_c)}\big)\dd{p^m}{z},
\label{segregation_force_multiPhase_model_appendix}
\end{equation}
with 
\begin{equation}
\mathcal{F}(\mu) = \big(1-e^{-70(\mu-\mu_c)}\big).
\label{segregation_function_multiPhase_model_appendix}
\end{equation}

\section{Link between the viscosity $\eta^p$ and the inertial number $I$}
\label{AppendixC}
The inertial number of large particles is 
\begin{equation}
I = \dfrac{\dot{\gamma}^m d_l}{\sqrt{p^m / \rho^p}}.
\end{equation}
Considering that
\begin{equation}
\eta^p = \dfrac{\tau^m_{xz}}{\dot{\gamma}^m} \quad \text{and} \quad  \tau^m_{xz} = \mu p^m,
\end{equation}
the inertial number can be written as
\begin{equation}
I = \dfrac{\mu d_l \sqrt{p^m \rho^p}}{\eta^p}.
\end{equation}
Then, making $\eta^p$ dimensionless with the scaling $\rho^p d_l W$, one can obtain
\begin{equation}
I = \dfrac{\mu \sqrt{\tilde{p}^m}}{\tilde{\eta^p}}
\end{equation}

\section{Deriving of the parameters $\mathcal{F}$ and $c$ using DEM results}
\label{AppendixD}
In this appendix, the methods to derive the advection and diffusion coefficients from the DEM simulations are developed. For the diffusion coefficient, it consists in proposing a new formulation of the drag coefficient, based on the small particle concentration, to found the accurate values of the diffusion coefficient. For the advection coefficient, it consists in rewriting the empirical segregation function $\mathcal{F}$, in order to find the same dependency with depth as the advection coefficient from the DEM simulations. 

\subsection{New formulation of the drag coefficient $c$}
In this part, the idea is to propose a new drag coefficient $c$ that satisfies
\begin{equation}
\dfrac{ \phi^s \sqrt{\tilde{p}^m} I}{\Phi 6 c \mu} = D_0I^{0.85}.
\label{dif_appendix}
\end{equation} 
Since it was found that
$$\displaystyle \dfrac{\sqrt{\tilde{p}^m} I}{\mu} = D_0^{\prime} e^{\tilde{z}/s_3} \quad \quad\text{with} \quad D_0^{\prime} = 8.46 \times 10^{-7} \quad  s_3 = 0.78 $$
and
$$\displaystyle I^{0.85} = I_0^{\prime} e^{\tilde{z}/s_2} \quad \quad\text{with} \quad I_0^{\prime} = 7.52 \times 10^{-7} \quad  s_2 = 0.8, $$
it can be written that
\begin{equation}
\dfrac{\sqrt{\tilde{p}^m} I}{ \mu I^{0.85}} = \dfrac{D_0^{\prime}}{I_0^{\prime}} = C_0,
\end{equation} 
where $C_0 = 1.13$.
In this way, equation \eqref{dif_appendix} becomes
\begin{equation}
c(\phi^s) = \phi^s \dfrac{C_0}{6 \Phi D_0} = 31 \phi^s
\label{}
\end{equation}

\subsection{New formulation of the empirical segregation function $\mathcal{F}$}
The new empirical segregation function has to fill the gap with the advection coefficient found using DEM. Therefore, $\mathcal{F}$ is such that:
\begin{equation}
\dfrac{I \mathcal{F}}{6 c(\phi^s) \mu \sqrt{\tilde{p}^m}} \vert\dd{\tilde{p}^m}{\tilde{z}}\vert =S_{r0}I^{0.85}.
\label{adv_appendix}
\end{equation}
In equation~\eqref{adv_appendix}, the dimensionless pressure gradient is constant and, using the hydrostatic approximation, reads
\begin{equation}
\partial \tilde{p}^m/ \partial \tilde{z} = - \Phi \left(\rho^p - \rho^f\right) / \rho^p.
\label{dimensionless_pressure}
\end{equation}
Introducing it in \eqref{adv_appendix}, it implies that the empirical segregation function must satisfies
\begin{equation}
\mathcal{F}(\mu, \tilde{p}^m, \phi^s) = \dfrac{6 S_{r0} c(\phi^s) \rho^p}{\Phi (\rho^p - \rho^f)}\mu \sqrt{\tilde{p}^m} I^{-0.15}.
\label{F_1_appendix}
\end{equation}

\bibliography{Grain_size_segregation}
\bibliographystyle{jfm}
\end{document}